\documentclass[%
 reprint,
superscriptaddress,
 amsmath,amssymb,
 aps,
 prb,
 longbibliography,
]{revtex4-2}

\usepackage{simplewick}
\usepackage{verbatim}
\usepackage{tikz}
\usetikzlibrary{shadings}
\usepackage{mathrsfs}
\usepackage{pgfplots}
\pgfplotsset{compat=1.17}
\usepackage{amsmath} 
\usepackage{graphicx}% Include figure files
\usepackage{xcolor}
\usepackage{pgfplots}
\usepgfplotslibrary{groupplots}
\usepackage{tikz}
\usepackage{pgfplotstable}
\pgfplotsset{compat=1.18}
\usepackage{mdframed}

\usepackage{dcolumn}% Align table columns on decimal point
\usepackage{bm}% bold math
\usepackage{physics}
\usepackage{subfigure}
\usetikzlibrary{decorations.markings, arrows.meta, patterns}
\usepackage[colorlinks=true, linkcolor=red, citecolor=blue, urlcolor=blue]{hyperref}

\begin{document}

\title{Entropy transport through a superfluid quantum point contact:\\[0.25 cm] A Keldysh field-theory approach}% Force line breaks with \\

%\author{Davide Bertolusso\textsuperscript{1}, C.J. Bolech\textsuperscript{1,2}, Thierry Giamarchi\textsuperscript{1}\\[0.20cm]
%\textsuperscript{1} DQMP, University of Geneva, 24 Quai Ernest-Ansermet, CH-1211 Geneva, Switzerland \\ 
%\textsuperscript{2} Department of Physics, University of Cincinnati, Ohio 45221, USA}

\author{Davide Bertolusso}
\email{davide.bertolusso@unige.ch}
\affiliation{DQMP, University of Geneva, 24 Quai Ernest-Ansermet, CH-1211 Geneva, Switzerland}
\author{C.J. Bolech}
\affiliation{DQMP, University of Geneva, 24 Quai Ernest-Ansermet, CH-1211 Geneva, Switzerland}
\affiliation{Dpt.\,of Physics, University of Cincinnati, 345 Clifton Court, OH-45221 Cincinnati, USA}
\author{Thierry Giamarchi}
\affiliation{DQMP, University of Geneva, 24 Quai Ernest-Ansermet, CH-1211 Geneva, Switzerland}

\date{\today}% It is always \today, today,
             %  but any date may be explicitly specified

\begin{abstract}

We study the matter and entropy transport between two ultra-cold neutral Fermi-gas reservoirs linked by a quantum point contact under a chemical-potential gradient. We describe the two leads with a BCS mean-field model and derive the current-bias characteristics for both particle and entropy transport. We compute the out of equilibrium steady-state currents by using the Keldysh formalism. In accordance with previous works in the literature, we confirm the well-known behavior for the particle current and extend the computation to the entropy current in the BCS regime. The entropy current shows an oscillatory behavior at low voltage in the ballistic junction limit. We analyze the results for a wide range of values of the junction's transparency. We also compare our findings with experimental results in cold atomic gases in the unitary regime.

\end{abstract}

%\keywords{Suggested keywords}%Use showkeys class option if keyword
                              %display desired
\maketitle

%\tableofcontents

\section{Introduction}
Recent advances in quantum-manipulation techniques for controlling individual atoms have enabled the experimental realization of a wide range of transport phenomena in both ultra-cold atomic systems and solid-state platforms. Foci of interest are transport of particles \cite{Valtolina_2015, Krinner_2014, Morpurgo, Del_pace_Tunneling, Husmann_2015, doi:10.1073/pnas.1601812113}, heat \cite{Brantut_2013} and entropy \cite{Fabritius_2024}. In particular, experiments with ultracold gases have allowed ballistic probing experiments \cite{Husmann_2015}, unlike in condensed matter where high-transparency tunneling is still difficult to achieve, due to the elastic scattering with phonons created by the lattice vibrations. Technological advances have also made it possible to control various aspects of a cold-atom system, such as dissipation \cite{Corman_2019}. The latter highlights the importance of cold-atom experiments as an excellent quantum platform for exploring novel physical phenomena. Indeed, the unexpected current behavior that has emerged from these recent works \cite{mohan2024universalentropytransportfar} has prompted the theoretical community to provide microscopic descriptions.
In addition, the ability to conduct a wider range of experiments stimulates the theoretical community because it allows for the study of different transport quantities \textcolor{black}{\cite{ PhysRevResearch.4.043087,PhysRevB.90.014511, Plastovets_2026}}, such as entropy flow. This gave us a more direct and broader access to ultra-cold physics. Particularly, out-of-equilibrium systems are a large subset of these experiments.

By implementing a tunneling-Hamiltonian approach \cite{Cuevas_1996}, we analyze a system of two superfluid reservoirs in separate local equilibria and connected by a point contact. The application of a gradient of chemical potential or temperature brings the system to an out-of-equilibrium configuration. Computational difficulties are associated with the consideration of a ballistic channel, \textit{i.e.}, tunneling with high transparency. The latter means that we need to consider higher-order processes and does not allow a direct perturbative treatment. From physical-relevance point of view, all these higher-order processes are of heightened interest within the superconducting gap; that is, when the chemical potential gradient is less than twice the superconducting gap $\Delta\mu<2\Delta$. Indeed, due to multiple Andreev reflection (MAR) process \cite{klapwijk1982} a particle from the lower band on the left reservoir can tunnel to the right reservoir by creating a multiple number of virtual Cooper pairs ($n_{pair}\geq2\Delta/2\Delta\mu$) \cite{Cuevas_2003}. This behavior is more evident for high-transparency tunneling; indeed, for low-transparency the currents, both particle and entropy, are suppressed inside the gap. Even if the limit of low transparency is well-known, the entropy current changes induced by increased transparency is still a largely unexplored question from a theoretical point of view. 

In this study, we investigate a model consisting of two superfluid reservoirs coupled via a quantum point contact (QPC), with the two subsystems described within the framework of standard BCS mean-field theory. The latter is reasonable, considering that such theoretical modeling perfectly reproduces the particle current as measured in the experiments \cite{Bolech_2005, Visuri_2023, Fabritius_2024}. The rest of the paper is organized as follows: In Sec.~\ref{Model_for_the_QPC} we introduce the model within the tunneling-Hamiltonian formalism and derive the entropy current. Then, in Sec.~\ref{Keldysh_formalism} we briefly explain the Keldysh approach needed to deal with out-of-equilibrium systems. In the final part, we show the results for the high (in Sec.~\ref{High}) and low (in Sec.~\ref{Lower}) transparency regimes, with a more detailed focus given to the ballistic case. Throughout the study we use the standard natural units, $\hbar=c=k_B=1$, and measure energies in units of the superfluid gap, $\Delta$.

\section{Model of the QPC}
\label{Model_for_the_QPC}
In this work, we consider a constriction linking two reservoirs, whose width is comparable to the Fermi wavelength and whose length is shorter than the superconducting correlation length \cite{Levy_Yeyati_1995}. Under these conditions, we can faithfully model the system through a quantum point contact. \textcolor{black}{To define the Hamiltonian, we first consider two isolated reservoirs with different particle numbers, i.e.\ different chemical potentials, while aligning the bottoms of the two bands. We then introduce the tunneling term and perform a gauge transformation $\psi_i \rightarrow \psi_i e^{-i \mu_i t}$ for later convenience in the calculations. The total Hamiltonian thus reads,}
\begin{comment}
\textcolor{red}{We use the grand-canonical ensemble which leads us to the following Hamiltonian given in the interaction picture \cite{mahan_many-particle_2000, Furutani2020}} \end{comment},
\begin{equation}
\label{eq:Hamiltonian}
\begin{aligned}
        H &= H_c + H_d + H_T = \\
           &\quad \sum_{\sigma,\mathbf{k},i \in\{c,d\}} 
             \bar{\psi}_{i}(t)_{\sigma,\mathbf{k}}
             \left(\epsilon_{i,\mathbf{k}} - \mu_i\right)
             \psi_{i}(t)_{\sigma,\mathbf{k}} \\ 
         &\quad
             + \sum_{\mathbf{k}, i \in\{c,d\}}
             \left[ \Delta_{i,\mathbf{k}}\,
             \bar{\psi}_{i}(t)_{\uparrow,\mathbf{k}}\,
             \bar{\psi}_{i}(t)_{\downarrow,-\mathbf{k}}
              + \text{h.c.}\right] \\ 
         &\quad
             - \tau \sum_{\sigma=\{\uparrow,\downarrow\},\mathbf{k},\mathbf{k}'} 
             \left[
             e^{i\Delta \mu t}\,
             \bar{\psi}_{c}(t)_{\sigma,\mathbf{k}} 
             \psi_{d}(t)_{\sigma,\mathbf{k}'} + \text{h.c.}
             \right]
\end{aligned}
\end{equation}
where $\epsilon_{i,\textbf{k}}$ is the energy dispersion for the left $(c)$ and right $(d)$ particles. 
In here we go beyond a linear-dispersion approximation around the Fermi energy, ($|\epsilon_{i,\textbf{k}}| = v_F |\textbf{k} - \textbf{k}_F|$, where $v_F$ is the Fermi velocity and $\textbf{k}_F$ the Fermi momentum), and used instead a more experimentally realistic quadratic dispersion, 
$\epsilon_{i,\textbf{k}} = \frac{\textbf{k}^2}{2m} - E_F$, with $m$ the mass of the fermionic particle, and $E_F$ the Fermi energy. We also define
$\mu_i$ as the chemical potential and $\Delta \mu = \mu_c - \mu_d$ as the chemical potential difference that enter as shifts in the particle distributions or as an explicit time dependence of the tunneling term. 
$\Delta_{i,\textbf{k}} = \Delta^*_{i,\textbf{k}} = \Delta_{i}$ 
is the superconducting gap and $\tau$ the tunneling-amplitude constant.
We work in a Nambu-spinor representation which leads us to the more compact Hamiltonian,

\begin{equation}
    H_i = \sum_{\textbf{k}} 
    \bar{\Psi}_{i,\textbf{k}}(t) 
    \big[(\epsilon_{i,\textbf{k}} - \mu_i)\sigma_z + \Delta\sigma_x\big]
    \Psi_{i,\textbf{k}}(t)
\end{equation}
and the tunneling term,

$$
H_T = 
- \tau \sum_{\textbf{k},\textbf{k}'} 
e^{i\Delta\mu t}
\bar{\Psi}_{c,\textbf{k}}(t)\, 
\sigma_z\,
\Psi_{d,\textbf{k}'}(t)
+ \text{h.c.}
$$
where the Nambu spinor is,
$$
\bar{\Psi}_{i,\textbf{k}}(t) = 
\left( 
\bar{\psi}_{i,\textbf{k},\uparrow}(t),\ 
\psi_{i,-\textbf{k},\downarrow}(t) 
\right)
$$
and $\sigma_{x,z}$ are the usual Pauli matrices \cite{shankar1994principles}. We model current's flow by a single-point tunneling from two separate bulks without boundaries.
%%% Notice PhysRev might not accept tikzpictures and one might have to generate and insert the corresponding PDF instead.
\begin{figure}[h!]
\centering
    \hspace{-0.3cm}\includegraphics[width=0.47\textwidth]{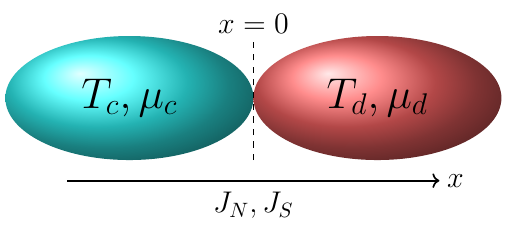}
    \begin{comment}
    \begin{tikzpicture}[scale=0.7]
    
    % Left and Right Blue Ellipses
    \shade[ball color=cyan!80!white] (-3,0) ellipse (3 and 1.5);
    \shade[ball color=red!60!white] (3,0) ellipse (3 and 1.5);
    
    % Labels L and R
    \node at (-3,0) {\huge \textbf{$T_c, \mu_c$}};
    \node at (3,0) {\huge \textbf{$T_d, \mu_d$}};
    
    \draw[dashed, thick] (0,-1.5) -- (0,1.5);
    \node at (0,1.8) {\Large $x=0$};
    
    % Horizontal axis and labels
    \draw[->, thick] (-4.5,-2) -- (4.5,-2) node[right] {\Large $x$};
    \node[below] at (0,-2.1) {\Large $J_N, J_S$};
    
    \end{tikzpicture}
    \end{comment}

\caption{Illustration of a two-reservoir system coupled at a single point $x=0$, filled with an $s$-wave superfluid, where the reservoirs may have distinct temperatures and chemical potentials. This graphical representation shows the particle, $J_N$, and entropy, $J_S$, currents flowing along the x-direction.}
\label{fig:nonequilibrium_reservoirs}

\end{figure} 

\subsection{Charge and Entropy current}
In order to extract the particle and entropy currents, we apply the Heisenberg equation of motion \cite{shankar1994principles},
\begin{equation}
    \frac{d}{d t} A(t)=i\left[H(t), A(t)\right]
\end{equation}
 to the particle number and heat operators, necessitating the evaluation of the associated commutators,
\begin{widetext}
\begin{equation}
\begin{aligned}
 J^{(c)}_N(t) 
&=\left\langle\frac{\mathrm{d} N_c}{\mathrm{d} t}\right\rangle
= i \left\langle 
\left[
H,\ 
\sum_{\textbf{k}, \sigma} 
\bar{\psi}_{c}(t)_{\sigma,\textbf{k}}\, 
\psi_{c}(t)_{\sigma,\textbf{k}}
\right]
\right\rangle 
\\
 J^{(c)}_{H}(t) 
&=\left\langle\frac{\mathrm{d} H_c}{\mathrm{d} t}\right\rangle
= i \left\langle 
\left[
H,\ 
\sum_{\textbf{k},\sigma} 
(\epsilon_{c,\textbf{k}} - \mu_c)\,
\bar{\psi}_{c}(t)_{\sigma,\textbf{k}}\, 
\psi_{c}(t)_{\sigma,\textbf{k}} + \sum_{\textbf{k}}
             \Delta_{c}\,
             \psi_{c}(t)_{\uparrow,\mathbf{k}}\,
             \psi_{c}(t)_{\downarrow,-\mathbf{k}}
              + \text{h.c.}
\right]
\right\rangle.
\end{aligned}
\end{equation}
Where for $\Delta_c \rightarrow 0$ we recover the current for the metallic limit.
Given by the two expressions above, we find the change of rate of particle and heat on the left reservoir. The corresponding quantities for the right reservoir are obtained by exchanging the two reservoirs. To show a direct link between the particle and heat currents, we evaluate the left and the right operators at two different times, leading to the following change of rates, first for the particle,
\begin{equation}
\label{eq:particle_current}
     J^{(c)}_N(t',t) 
    = -i \left\langle \sum_{\textbf{k},\textbf{k}',\sigma} 
    \tau \Big[
        e^{i\Delta \mu t'} 
        \bar{\psi}_{c}(t)_{\sigma,\textbf{k}}\,
        \psi_{d}(t')_{\sigma,\textbf{k}'}
        - \text{h.c.}
    \Big] \right\rangle
\end{equation}
then for heat as,
\begin{equation}
\label{eq:heat_current}
     J^{(c)}_H(t',t) 
    = - \left\langle \sum_{\textbf{k},\textbf{k}',\sigma} 
    \tau \Big[
        e^{i\Delta \mu t'} 
        \Dot{\bar{\psi}}_{c}(t)_{\sigma,\textbf{k}}\,
        \psi_{d}(t')_{\sigma,\textbf{k}'}
        + \text{h.c.} 
    \Big] \right\rangle 
\end{equation}
\end{widetext}
where $\Dot{\bar{\psi}}_{c}(t)_{\sigma,\textbf{k}} = i\left[H(t), \bar{\psi}_{c}(t)_{\sigma,\textbf{k}}\right]$. \textcolor{black}{A brief derivation of the above currents operator is describe in Appendix \ref{Currents}}. The form of the heat rate given in Eq.~\eqref{eq:heat_current} is particularly convenient \cite{Uchino_2021}, as it facilitates a direct connection to the two-body operator in Eq.\eqref{eq:particle_current},
\begin{equation}
    \Dot{\bar{\psi}}_{c}(t)_{\uparrow,\textbf{k}}\,
        \psi_{d}(t')_{\uparrow,\textbf{k}'} = \partial_{t} \bar{\psi}_{c}(t)_{\uparrow,\textbf{k}}\,
        \psi_{d}(t')_{\uparrow,\textbf{k}'}.
\end{equation}
In frequency space we get the following two-body operators for the rate of heat transfer,
\begin{equation}
\label{current_FT}
\begin{aligned}
    \textcolor{black}{e^{i\Delta \mu t'}} \Dot{\bar{\psi}}_{c}(t)_{\sigma,\textbf{k}}\,
        \psi_{d}(t')_{\sigma,\textbf{k}'} &= \int d\omega d \omega' i\omega \bar{\psi}_{c}(\omega)_{\sigma,\textbf{k}}\,
        \psi_{d}(\omega')_{\sigma,\textbf{k}'} \\ 
        &\quad \times e^{i\omega t}e^{-i\omega' t'}\textcolor{black}{e^{i\Delta \mu t'}}  \\ 
        \textcolor{black}{e^{-i\Delta \mu t'}}\bar{\psi}_{d}(t')_{\sigma,\textbf{k}'}\, \Dot{\psi}_{c}(t)_{\sigma,\textbf{k}} &= -\int d\omega d \omega' i\omega \bar{\psi}_{d}(\omega')_{\sigma,\textbf{k}'}\,
        \psi_{c}(\omega)_{\sigma,\textbf{k}} \\ 
        &\quad \times e^{i\omega' t'}e^{-i\omega t}\textcolor{black}{e^{-i\Delta \mu t'}} .
\end{aligned}
\end{equation}
Since in this work we are only focusing on the DC component, after the limit $t \rightarrow t'$ is taken we average out the AC component by using the following expression for the Dirac delta function,
\begin{equation}
    \lim_{T \rightarrow \infty} \frac{1}{T} \int_{-T/2}^{T/2} e^{\pm i (\omega - \omega' \textcolor{black}{+ \Delta \mu})t} dt = \delta(\omega - \omega'  \textcolor{black}{+ \Delta \mu}).
\end{equation}
Finally, with the chemical potential term included, the rates of particle and heat transfer are given by,
\begin{equation}
\begin{gathered}
    \label{eq:current_final_super_nokeldysh}
     J^{(c)}_n  = \frac{\tau}{4\pi} \sum_{\sigma} \int \omega^n 2\Re [-i \langle \bar{\psi}_{c,\sigma}(\omega) \psi_{d,\sigma}(\omega+\Delta \mu) \rangle]  d\omega
\end{gathered}
\end{equation}
where $n=0$ gives the particle (N), while $n=1$ is the heat (H), and we have summed over momentum space.
Therefore, the central quantity to evaluate is the correlation between the left and right fields.

The total current of interest is given by the difference between the change of rate of particle and heat associated with the left reservoir, $J_n^{(c)}$, and those associated with the right reservoir, $J_n^{(d)}$,
\begin{equation}
\label{eq:current_total}
    J_n = -\frac{ J_n^{(c)}  -   J_n^{(d)} }{2}
\end{equation}
as detailed in Appendix~\ref{Thermodynamics}.
Up to this point, we have focused on the heat current. However, we ultimately present the entropy current, which can be easily related to the heat current, through thermodynamics considerations, as explained also in Appendix~\ref{Thermodynamics},
    \begin{equation}
\begin{aligned}
    J_S &= \frac{1}{4T^2 - \Delta T^2}\left( 4T J_U  - (4\mu T - \Delta \mu \Delta T)J_N\right) \\
    &= \frac{1}{4T^2 - \Delta T^2}\left( 4TJ_H + \Delta \mu \Delta TJ_N\right)
\end{aligned}
\end{equation}
where $J_U$ is the energy current, computed within the canonical ensemble.

In the next section we summarize the Keldysh path integral formalism that is essential to compute the integrand kernels needed in an out-of-equilibrium setting, and consequently the particle and entropy currents. 
\begin{figure}[t]
    \centering
    \includegraphics[width=0.3\textwidth]{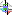} % Replace with your image file name
    \caption{Depiction of a \emph{Andreev composite tunneling} process at zero temperature ($T_c = T_d = 0$) by involving two virtual Cooper pairs. Here, the chemical potential drop enters as a shift rather than as an explicit time dependence in the tunneling as in Eq.\eqref{eq:Hamiltonian}, and it is numerically set at $\Delta \mu =1.2\Delta$. The light -green process represents a particle from the left reservoir that tunnels elastically to the right while entangled with two virtual Cooper pairs, one on each lead (or via two successive Andreev reflections). The light-blue is the dual hole process starting from the right reservoir. Black (white) circles represent spin-up particles (spin-down holes) and on the vertical axis we give the single-particle density of states of each lead (with the shading representing the zero-temperature occupation). The above process occurs by $n=n_{\text{pair}}+1=3$ ``cotunneling'' events, where each of them contributes with a transmission of $\mathcal{T}^2$, therefore giving a total transmission weight of $\mathcal{T}^6$. The pictured process can occur when $ \Delta\mu > 2/3 \Delta$.}
    \label{fig:Andreev_process}
\end{figure}
\begin{figure}[t]
    \centering
    \includegraphics[width=0.3\textwidth]{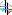} % Replace with your image file name
    \caption{Depiction of a \emph{single Andreev reflection} process at zero temperature ($T_c = T_d = 0$). In this representation, the chemical potential enters as a shift rather than an explicit time dependence in the tunneling term as in Eq.\eqref{eq:Hamiltonian}, and it is numerically set at $\Delta \mu =1.2\Delta$, as in Fig.~\ref{fig:Andreev_process}. The light-green process represents a particle from the left reservoir that is being reflected as a hole to the same reservoir via one virtual Cooper pair on the right reservoir. The light-blue instead is the dual hole representation starting from the right reservoir. Black (white) circles represent spin-up particles (spin-down holes). The above process occurs with a transmission weight of $\mathcal{T}^4$, given that the transport is happening through two ``cotunneling'' events. This process can occur only if $\Delta\mu > \Delta$.}
    \label{fig:Andreev_process_reflection}
\end{figure}

\section{Keldysh formalism}
\label{Keldysh_formalism}

\begin{figure}[t]
% TikZ figures might need to be given in PDF format...
    \centering
    \includegraphics[width=0.4\textwidth]{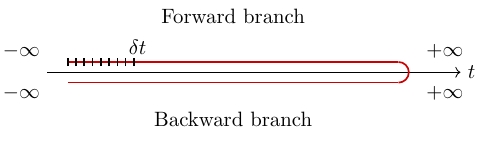}
    \begin{comment}
    \begin{tikzpicture}[scale=0.7]
    % Time axis
    \draw[->] (-4,0) -- (6,0) node[right] {$t$};

    % Keldysh contour
    \draw[thick, red!80!black] (-3.5,-0.25) -- (0.5,-0.25);
    \draw[thick, red!80!black] (0.5,-0.25) -- (4.5,-0.25);
    \draw[thick, red!80!black] (4.5,-0.25) arc(-90:90:0.25);
    \draw[thick, red!80!black] (-3.5,0.25) -- (0.5,0.25);
    \draw[thick, red!80!black] (0.5,0.25) -- (4.5,0.25);

    \draw[thick] (-2.5,0.15) -- (-2.5,0.35);
    \draw[thick] (-2.3,0.15) -- (-2.3,0.35);
    \draw[thick] (-2.1,0.15) -- (-2.1,0.35);
    \draw[thick] (-1.9,0.15) -- (-1.9,0.35);
    \draw[thick] (-2.7,0.15) -- (-2.7,0.35);
    \draw[thick] (-2.9,0.15) -- (-2.9,0.35);
    \draw[thick] (-3.1,0.15) -- (-3.1,0.35);
    \draw[thick] (-3.3,0.15) -- (-3.3,0.35);
    \draw[thick] (-3.5,0.15) -- (-3.5,0.35);
    
    % Labels for -∞ and +∞
    \node at (-4,-0.5) [left] {$-\infty$};
    \node at (-4,0.5) [left] {$-\infty$};
    \node at (5,-0.5) [right] {$+\infty$};
    \node at (5,0.5) [right] {$+\infty$};

    % Delta t arrow
    \node at (-2.2, 0.6) [right] {$\delta t$};

    \node at (0.5, 1) [above] {Forward branch};
    \node at (0.5,-1.5) [above] {Backward branch};

    \end{tikzpicture}
    \end{comment}
    
\caption{Graphical representation of the non-equilibrium closed-path Keldysh integral. On the upper branch is where the \textit{forward} field lives; instead, the \textit{backward} field lives on the lower branch of the path.}
\label{fig:nonequilibrium_path}
\end{figure}

In order to compute the currents in an out-of-equlibrium system we need to implement the Keldysh formalism \cite{Kamenev_2011}. The main feature of the Keldysh formalism is the closed-path integral, which allows us to evaluate the required two-point functions; in particular, the integrand kernels of the currents. With respect to the usual open-path integral, the closed-path one requires two copies of the fields, one on the \textit{forward} branch, $\psi_+$, and one on the \textit{backward} branch, $\psi_-$, as graphically shown in Fig.~\ref{fig:nonequilibrium_path}. We then perform the Keldysh rotation of these fields,
\begin{equation*}
    \begin{pmatrix}
        \psi_+ \\
        \psi_-
    \end{pmatrix}
    \rightarrow
        \begin{pmatrix}
        \psi^{cl} \\
        \psi^{q}
    \end{pmatrix} = 
    \begin{pmatrix}
        \frac{1}{\sqrt{2}} (\psi_+ + \psi_-) \\
        \frac{1}{\sqrt{2}} (\psi_+ - \psi_-)
    \end{pmatrix}
\end{equation*}
where we introduce the \textbf{\textit{classical}} and \textbf{\textit{quantum}} fields, respectively. Regarding the Green functions, we have the following transformation \cite{Kamenev_2011},
\begin{equation*}
        \begin{pmatrix}
        G_{++}  & G_{+-}\\
        G_{-+}   & G_{--}
    \end{pmatrix}
    \rightarrow
        \begin{pmatrix}
        G_K  & G_R\\
        G_A   & 0 
    \end{pmatrix}
\end{equation*}
where,
$$
    G_{ab} = -i\langle \psi_a(t) \bar{\psi}_b (t') \rangle
$$
with $a,b = \pm$. The zero entry of the latter matrix is a direct consequence of the density-matrix conservation law \cite{Sieberer_2016}.

Starting from the definition of the mean value of an operator,
\begin{equation}
    \langle{O}\rangle = \int \mathcal{D}[\bar{\psi} \psi]\;O\,e^{i S[\bar{\psi}, \psi]}
\end{equation}
we can compute the physical quantities of interest, such as the particle and entropy currents; but we must first define the action corresponding to our system. The total action, $S$, is the sum of the actions of the left and right reservoirs, $S_L$ and $S_R$, plus the tunneling term $S_T$. Particularly, in real space these actions read as follows,
\begin{equation}
\begin{aligned}
    S_i &= \int dt \sum_\textbf{k}  \bar{\Psi}_{i,\textbf{k}}(t) [(\partial_t-\epsilon_{i,\textbf{k}} +\mu_i)\sigma_z - \Delta\sigma_x]\Psi_{i,\textbf{k}}(t) \\
    S_T &= \tau\int dt\sum_\textbf{k}  e^{i\Delta\mu\,t}\bar{\Psi}_{i,\textbf{k}}(t) \sigma_z \Psi_{i,\textbf{k}}(t) + \text{h.c.}
\end{aligned}
\end{equation}
The structure of the action is nontrivial, as the superconducting gap couples frequencies that are symmetric with respect to the chemical potential, while the tunneling term connects states at the same ($\Delta\mu$-shifted) frequencies across the leads. This interplay significantly complicates the analysis. 
In Keldysh space, the rate of particle and heat transfer reads as,
\begin{equation}
\label{eq:current_final_super}
     J^{(c)}_n  = \frac{\tau}{4\pi} \sum_{\sigma} \int d\omega \omega^n 2\Re [-i\langle \bar{\psi}_{c,\sigma}^{cl}(\omega) \psi_{d,\sigma}^{cl}(\omega+\Delta \mu)\rangle].
\end{equation}
We omit the contributions involving the \textbf{\textit{quantum}} fields, as they all vanish identically for off-diagonal terms. Consequently, the only relevant nonzero two-point term is the \textbf{\textit{classical--classical}} correlation.

In the following section, we want to better understand the structure of the action in the Keldysh basis for the two uncoupled leads and the tunneling term in frenquency-momentum space.

\subsection{Keldysh Green functions}
\label{Keldysh Green functions}
In this section, we derive the Keldysh Green functions for our system, considering both a quadratic dispersion relation and a superconducting BCS system. We begin by writing the \textit{local} action of the uncoupled leads,
\begin{equation}
    S_i = \int dt\; 
    \bar{\Psi}(\textbf{r}=\mathbf{0},t)\,
    \mathcal{G}^{-1}(\textbf{r}=\mathbf{0},t)\,
    \Psi(\textbf{r}=\mathbf{0},t),
\end{equation}
where we integrate out all the points different from the point of tunneling contact.
The corresponding inverse Green functions in Fourier space naturally follow as,
\begin{equation}
\label{eq:green_action}
    \mathcal{G}^{-1}(\omega,\textbf{k}) 
    = 
    \begin{pmatrix}
        0 & G_A^{-1} \\
        G_R^{-1} & G_K^{-1}
    \end{pmatrix}(\omega,\textbf{k}) \,.
\end{equation}
Above, the components of the Green functions can be either scalar functions, as for the case of a metal, or matrices in Nambu space, as for the case of a superconductor.
In what follows, we briefly report the components of the green function for the metallic case first and after we give the result for the more involved superconducting case. 

\subsubsection{Metal}

For metallic leads, the components of the Green function in momentum and frequency space are 
\cite{Kamenev_2011},
\newline
\begin{equation}
\begin{aligned}
        G^{R}(\omega,\textbf{k}) &= 
        \lim_{\eta \to 0}
        \frac{1}{\omega - (\epsilon_{\textbf{k}} - \mu) + i\eta}, \\
        G^{A}(\omega,\textbf{k}) &= 
        \lim_{\eta \to 0}
        \frac{1}{\omega - (\epsilon_{\textbf{k}} - \mu) - i\eta}, \\
        G^{K}(\omega,\textbf{k}) &= 
        \tanh\!\left(\frac{\omega}{2T}\right)
        (-2\pi i)\, \delta((\epsilon_{\textbf{k}} - \mu) - \omega).
\end{aligned}
\end{equation}
where $\eta$ is a positive infinitesimal introduced to regularize the expression and physically representing an inelastic scattering rate and $T$ and $\mu$ are respectively the temperature and the chemical potential of the lead. The above Green functions are, respectively, the retarded, advanced, and Keldysh components. The latter follows directly from the fluctuation--dissipation theorem, as each of the two reservoirs is assumed to be in thermal equilibrium \cite{Kamenev_2011}.  
Since we focus on a single tunneling point contact, we integrate over momentum to obtain the local retarded Green function,
\begin{widetext}
\begin{equation}
\label{eq:green_retarded_metal}
\sum_{|\mathbf{\textbf{k}}| \le k_{\max}} G^R(\omega,k)
=
\frac{2 m^{3/2}}{2 \pi^2}\left[- k_{\max}
+
\sqrt{-E_F-\mu-(\omega + i \eta)}\;
\arctan\!\left(
\frac{k_{\max}}{\sqrt{-E_F-\mu-(\omega + i \eta)}}
\right) \right]
\end{equation}
\end{widetext}
where we introduce an ultraviolet cutoff (UV) in momentum space, denoted by $k_{\max}$, such that the summation over momentum space is restricted according to
\begin{equation}
\sum_{\mathbf{k}} \;\longrightarrow\; \sum_{|\mathbf{k}| \le k_{\max}} \,
\end{equation}
The corresponding advanced and Keldysh Green functions then follow straightforwardly from the retarded component \cite{Kamenev_2011}. Finally, the Green function is inverted to extract the relevant matrix element of the action, as described in Eq.~\eqref{eq:green_action}.

\subsubsection{Superconductor}

In the superconducting case, the Green function acquires a non-trivial matrix structure. Besides the usual normal components associated with spin-up and spin-down sectors, the Nambu spinor formalism introduces the anomalous off-diagonal terms, which capture the spin-pair correlations characteristic of the BCS state,
\begin{widetext}
\begin{equation}
\begin{gathered}
    G^{R/A}(\omega, \textbf{k})
    =
    \frac{1}{
        (\omega \pm i\eta)^2
        - (\epsilon_{\textbf{k}} - \mu)^2
        - \Delta^2
    }
    \begin{pmatrix}
        (\omega \pm i\eta) + (\epsilon_{\textbf{k}} - \mu) 
        & -\Delta \\
        -\Delta 
        & (\omega \pm i\eta) - (\epsilon_{\textbf{k}} - \mu)
    \end{pmatrix},
\end{gathered}
\end{equation}
\end{widetext}
The superscripts $R$ and $A$ correspond to the choices of the $\pm$ sign, respectively.  
In the following, we omit the explicit notation $\lim_{\eta\to 0}$, with the understanding that $\eta$ denotes a positive infinitesimal. In the following, we focus on the retarded Green function, since the advanced one is its conjugate \cite{Kamenev_2011}.
From the fluctuation dissipation theorem we write the Keldysh green function,
\begin{equation}
    G^{K}(\omega,\textbf{k}) 
    = 
    \tanh\!\left(\frac{\omega}{2T}\right)
    \left( 
        G^{R}(\omega,\textbf{k}) - 
        G^{A}(\omega,\textbf{k})
    \right).
\end{equation}
By using the free quadratic dispersion we obtain the following matrix expression for the inverse retarded Green function at the contact point for the superconductor,
\begin{equation}
\label{eq:superconducting_green}
    \left[ \sum_{\textbf{k}} G^{R}(\omega, \textbf{k}) \right] ^{-1} = \frac{2 \pi^2}{(2m)^{3/2}}\frac{1}{A_-A_+-A_\text{off}^2}\begin{bmatrix}
        A_- & -A_\text{off} \\
        -A_\text{off} & A_+
    \end{bmatrix} 
\end{equation}
where the $\pm$ index indicates the two Nambu components of the fermions. We define the different matrix components below.
Since our goal is to compute the entropy current, the linear-dispersion approximation is no longer sufficient, since particle-hole asymmetry must be incorporated into the model. This is naturally achieved by adopting a quadratic dispersion relation, which leads to the following expressions parameterizing the local Green functions,
\begin{widetext}
\begin{equation}
\begin{aligned}
A_+ &= -k_{\max} + \left[ F_{\uparrow}(f(\omega), \omega, k_{\max}, E_F+\mu, \eta) + F_{\uparrow}(-f(\omega), \omega, k_{\max}, E_F+\mu, \eta) \right] \\
A_- &= k_{\max} + \left[ F_{\downarrow}(f(\omega), \omega, k_{\max}, E_F+\mu, \eta) + F_{\downarrow}(-f(\omega), \omega, k_{\max}, E_F+\mu, \eta) \right]
\end{aligned}
\end{equation}
where we defined,
\begin{equation}
\begin{aligned}
F_{\uparrow}(f(\omega), \omega, k_{\max}, E_F+\mu, \eta) &= \frac{
\left( \Delta^2 - ( (E_F +\mu + \omega)+i\eta )\left(f(\omega) + \omega + i\eta \right)\right)
\;
\arctan\!\left[
    \frac{k_{\max}}{
        \sqrt{-E_F-\mu - f(\omega)}
    }
\right]
}{2f(\omega)\sqrt{-E_F - \mu - f(\omega)}} \\
F_{\downarrow}(f(\omega), \omega, k_{\max}, E_F+\mu, \eta) &= \frac{
\left( \Delta^2 - ( (-E_F-\mu + \omega)+i\eta )\left(-f(\omega) + \omega + i\eta \right)\right)
\;
\arctan\!\left[
    \frac{k_{\max}}{
        \sqrt{-E_F-\mu + f(\omega)}
    }
\right]
}{2f(\omega)\sqrt{-E_F - \mu + f(\omega)}}
\end{aligned}
\end{equation}
%\end{widetext}
with,
$f(\omega) = \sqrt{(\omega + i\eta)^2 - \Delta^2}$.
We further evaluate the off-diagonal component associated with the superfluid gap,
%\begin{widetext}
\begin{equation*}
\begin{aligned}
    A_\text{off} = \frac{-\Delta}{2 f(\omega)}\left(-\sqrt{-E_F-\mu-f(\omega)}\arctan\!\left[
    \frac{k_{\max}}{
        \sqrt{-E_F-\mu - f(\omega)}}\right] + \sqrt{-E_F-\mu+f(\omega)}\arctan\!\left[
    \frac{k_{\max}}{
        \sqrt{-E_F-\mu + f(\omega)
    }}
\right]\right)
\end{aligned}
\end{equation*}
%\end{widetext}
By taking the limit of the gap going to zero, we can recover the retarded Green function for the metallic case. Indeed, we can easily see the off-diagonal matrix element going to zero while, instead, the diagonal element becomes,
%\begin{widetext}
\begin{equation}
G^{R,-1}_{\uparrow \uparrow} = 
\frac{2 \pi^2}{2 m^{3/2}} \left[- k_{\max}
+
\sqrt{-E_F-\mu-(\omega + i \eta)}\;
\arctan\!\left(
\frac{k_{\max}}{\sqrt{-E_F-\mu-(\omega + i \eta)}}
\right) \right]^{-1}
\end{equation}
\end{widetext}
The above expression is consistent with the metallic case given by Eq.~\eqref{eq:green_retarded_metal}.
The overall system is not block diagonal in frequency and therefore, inspired by \cite{Visuri_2023} we can write the action in a clever matrix way, such that after a finite number of MARs the off-diagonal terms will tend to vanish (more details below).
\section{High transparency}
\label{High}
Throughout this section, we focus on the high-transparency regime, $\mathcal{T} \rightarrow 1$. Subsequently, we also present results corresponding to intermediate and low transparency. We begin by deriving the expression for the current in a Normal--Normal junction, followed by the corresponding analysis for a Superconductor--Superconductor junction. In addition, we briefly discuss the behavior of a Normal--Superconductor junction, which will be relevant for later considerations. Throughout the discussion, we use the parameters $E_F = 20\Delta$, $k_{\max} = 3E_F$, $T_c = T_d = T = 0.15\Delta$, $\mu=(\mu_c - \mu_d)/2=0$ and $\eta = 10^{-7}\Delta$. All results are normalized with respect to \(G_0 \Delta\), where \(G_0 = \frac{2}{h}\) denotes the quantum of conductance. We verified that taking the ultraviolet cutoff, $k_{\max}$, to larger values does not significantly change the results.

\subsection{Normal-Normal junction}
Before going into the case of superconducting leads we briefly review the results for two metallic reservoirs in the normal state, $\Delta_c = \Delta_d = 0$. Generally, without any assumption on the energy dispersion of the system, the particle and heat current read as follows,
\begin{equation}
\label{eq:current_formula_metal}
\begin{aligned}
    J_n = G_0 \int d\omega \omega^n \mathcal{T}^2(\omega) (n_c(\omega-\Delta \mu /2) - n_d(\omega+\Delta \mu /2))
\end{aligned}
\end{equation}
The above expression corresponds to the standard Landauer-Büttiker formula for transport between two non-interacting reservoirs \cite{Landauer_1957, Buttiker, Berthod-PRB-2011, Todorov_1993}. The functions $n_c$ and $n_d$ denote the Fermi-Dirac distribution functions of the left and right reservoirs, respectively. We define $\mathcal{T}(\omega)$ as the transmission function, given by,
\begin{equation}
\label{eq:metallic_transmission}
    \mathcal{T}^2(\omega) = \frac{4 \pi^2 \tau^2 \rho_c(\omega) \rho_d(\omega)}{\left| 1 - G^R_c(\omega) G^R_d(\omega) \tau^2 \right|^2}
\end{equation}
where $\rho_{c/d}(\omega) = -\frac{1}{\pi} \Im \big[ G^R_{c/d}(\mathbf{r}=\mathbf{0}, \omega) \big]$ is the local density of states (LDoS). Here, the functions with subscript $c$ are evaluated at $\omega - \Delta \mu/2$, while those with subscript $d$ are evaluated at $\omega + \Delta \mu/2$. The transmission function $\mathcal{T}(\omega)$ describes the probability for a particle to tunnel and represents the relevant physical parameter, as highlighted in Appendix~\ref{Transperancy}.

For a perfectly linear dispersion relation, the transmission function is a constant at it simplifies to,
\begin{equation}
    \mathcal{T}^2 = \frac{4 \pi^2 \alpha^2}{(1 + \pi^2 \alpha^2)^2}
\end{equation}
with $\alpha^2 = \rho_0^2 \tau^2$, assuming a constant density of states $\rho_0$ near the Fermi energy. In this case, the kernel of the current is odd for $n=1$, which leads to a vanishing entropy current and highlights the need for quadratic corrections.

We emphasize that the above formulas, in the metallic limit, satisfy the Onsager relations \cite{Brantut_2013}, reproduce the Seebeck coefficient, and, at low temperatures, are consistent with the Wiedemann-Franz law \cite{kittel2005introduction}.

\subsection{Superconductor-Superconductor junction}

\begin{figure}[t]
    \centering
    \hspace{-0.4cm}\includegraphics[width=0.47\textwidth]{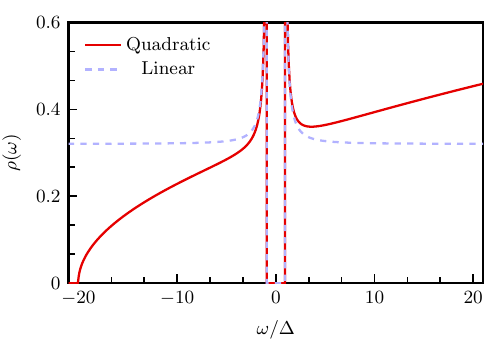}
    % Replace with your image file name
    \caption{The local density of states (LDoS) as a function of frequency, with $E_F = 20\Delta$ for the linear (dashed line) and quadratic dispersion (solid line).}
\label{fig:DOS}
\end{figure}

We now turn to the superconducting case. Here, for simplicity, the two superconducting gaps are taken to be equal, \(\Delta_c = \Delta_d = \Delta\). The  LDoS for the superconducting case is obtained,
\begin{equation}
    \rho(\omega) = -\frac{1}{\pi} \Im{G^{R}_{\uparrow \uparrow}(\textbf{r}=\mathbf{0}, \omega)}
\end{equation}
and we show in Fig.~\ref{fig:DOS} how it gets modified when a superconducting gap opens around the Fermi level. The linear approximation (dashed line) reproduces well the density near the gap edges but fails to capture the asymmetry due the overall curvature of the band, which is essential to model the heat and entropy currents.

For superconducting leads the system is no longer diagonal in the frequency domain; physically this is understood through multiple Andreev reflection (MARs), as was shown graphically in Fig.~\ref{fig:Andreev_process}.

Referring back to Eq.~\eqref{eq:current_final_super}, we need to compute the \textbf{\textit{classical--classical}} Keldysh correlation function. Once the expression for the particle current is obtained, the next step involves inverting the Green function appearing in the action of the full coupled system to extract the current's integrand. While the Green function is, in principle infinite-dimensional, the off-diagonal block elements decay, when computed with respect to the reservoir chemical potential, as $|\omega-\mu|$ grows, that can be seen from the off-diagonal component of the local Green function in Eq.~\eqref{eq:superconducting_green}, or more clearly from the linear dispersion approximation \cite{Visuri_2023}. The latter allows the truncation of the matrix after a finite number of Andreev reflections \cite{Bolech_triplet, Visuri_2023}.
\textcolor{black}{The detailed construction of the superconducting action is discussed in Appendix \ref{Action_structure}}. In our numerical analysis, we included up to $n\!=\!100$ ``co-tunneling events'' for the particle current, (cf.~with $n\!=\!3$ in Fig.~\ref{fig:Andreev_process}), and up to $n\!=\!50$ for the entropy current. The number of co-tunneling events is related to the number of Cooper-pair tunneling events, or MAR processes, through the relation $n_{\text{pair}}\!=\!n-1$.
Consequently, for bias values below $\Delta \mu\!=\!0.02\Delta$  multiple Andreev reflections are suppressed in the particle current, and below $\Delta\mu\!=\!0.04\Delta$ for the entropy current. We note that even when the action is truncated, the current is inherently suppressed since the transparency $\mathcal{T}$ is not exactly unity. Indeed, the current scales as $\mathcal{T}^{2n}$ and is thus already strongly suppressed before reaching the numerical MARs cutoff.
 
\begin{figure*}[t]
    \centering
    %\hspace{-1cm}\input{plots/099_SS_Particle}
    %\hspace{1cm}\input{plots/099_SS_Entropy}
    \centering  \hspace{-0.5cm}\includegraphics[width=1.02\textwidth]{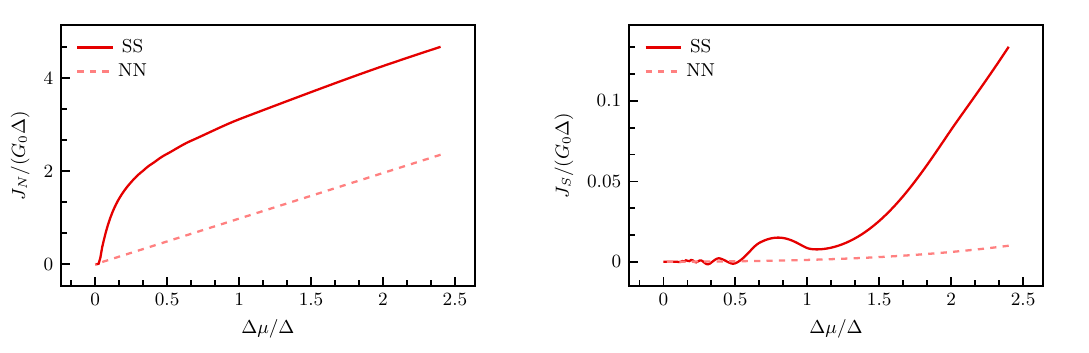}
%    \caption{Current-voltage behavior for the particle transport under a chemical potential bias at $\mathcal{T} = 0.99$. The solid line is the Superconducting-Superconducting (SS) junction, instead the dashed line is the Normal-Normal (NN) junction. The plot shows the non-linear behavior for the SS junction shown for comparison. We can see a suppression of the current at $\Delta \mu = 0.02 \Delta$ due to the finite number of MARs taken into account in the numerical analysis.}
    \caption{Current-voltage behavior for the particle (left panel) and entropy (right panel) transport under a chemical potential bias at $\mathcal{T} = 0.99$. The solid line is the Superconducting-Superconducting (SS) junction, instead the dashed line is the Normal-Normal (NN) junction shown for comparison. The plots show the non-linear behavior for the SS junction, displaying oscillations in the case of the entropy current. We can see a suppression of the current at $\Delta \mu = 0.02 \Delta$ ($\Delta \mu = 0.04 \Delta$ for the entropy current) due to the finite number of MARs taken into account in the numerical calculation. We emphasize that, even when including a higher number of MAR processes, the current will ultimately start to converge to zero around the same $\Delta \mu$ due to the non-unitary value of $\mathcal{T}$.} 
    \label{fig:Particle_Entropy_current_superconductor}
    \vspace{0.1cm}
    \centering  \hspace{-0.5cm}\includegraphics[width=1.02\textwidth]{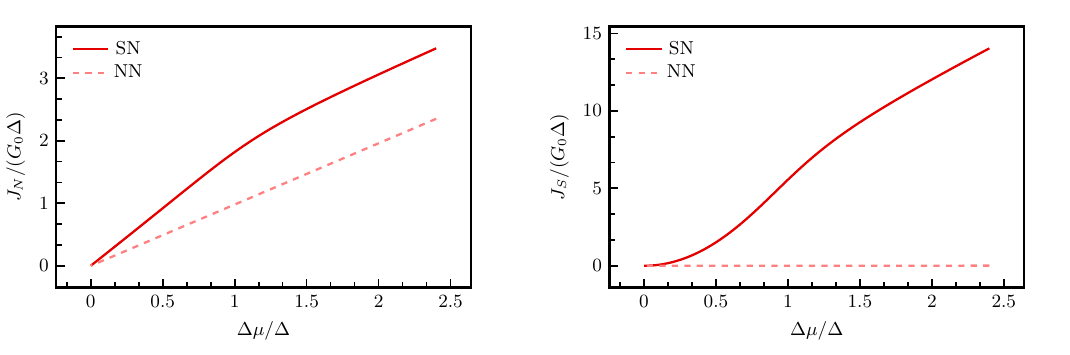}
    \caption{Particle and entropy currents as a function of voltage for a SN junction under an applied chemical potential bias at transparency $\mathcal{T}=0.99$. The figure isolates the contribution arising from a single Andreev reflection process in the subgap region and compares it with the Ohmic response of the normal limit, in agreement with \cite{blonder1982}. The right panel highlights the quadratic-like behavior of the entropy current in the subgap region. The metallic limit is shown as a dashed line.}
    \label{fig:Particle_Entropy_current_SN}
\end{figure*} 

The particle current is shown in left panel of Fig.~\ref{fig:Particle_Entropy_current_superconductor}, where we observe a clear enhancement in the current even at small chemical potential differences, consistent with the presence of a ballistic channel. This behavior aligns with earlier work \cite{Visuri_2023,Yao_2018}, and confirms the validity of the linear approximation for the energy dispersion. In contrast, the entropy current shows an oscillating behavior (see right panel of the same figure), which we will focus on in Section~\ref{Lower}. 

\subsection{Superconductor-Normal junction}

To gain a clearer understanding of the entropy current, we first examine the behavior of a Superconductor-Normal metal (SN) junction (see Fig.~\ref{fig:Particle_Entropy_current_SN}).
To that end, we impose $\Delta_c\!=\!\Delta$ and $\Delta_d\!=\!0$.

In this case, only a single particle-hole reflection occurs \cite{blonder1982}, with no contribution from multiple Andreev reflections, (see, e.g., Fig.~\ref{fig:Andreev_process_reflection}).
This allows us to isolate and clearly identify the entropy current arising from this process.
In particular, the plots in Fig.~\ref{fig:Particle_Entropy_current_SN} show the particle, in agreement with \cite{blonder1982}, and entropy currents, respectively, for an SN junction in the ballistic limit.
From the latter, let us highlight the quadratic behavior inside the gap (crossing over to linear outside). This observation will be relevant for later discussions. 

\section{Low transparency}
\label{Lower}

\begin{figure*}[t]
\centering
%\hspace{-1cm}
\hspace{-1cm}\includegraphics[width=1.00\textwidth]{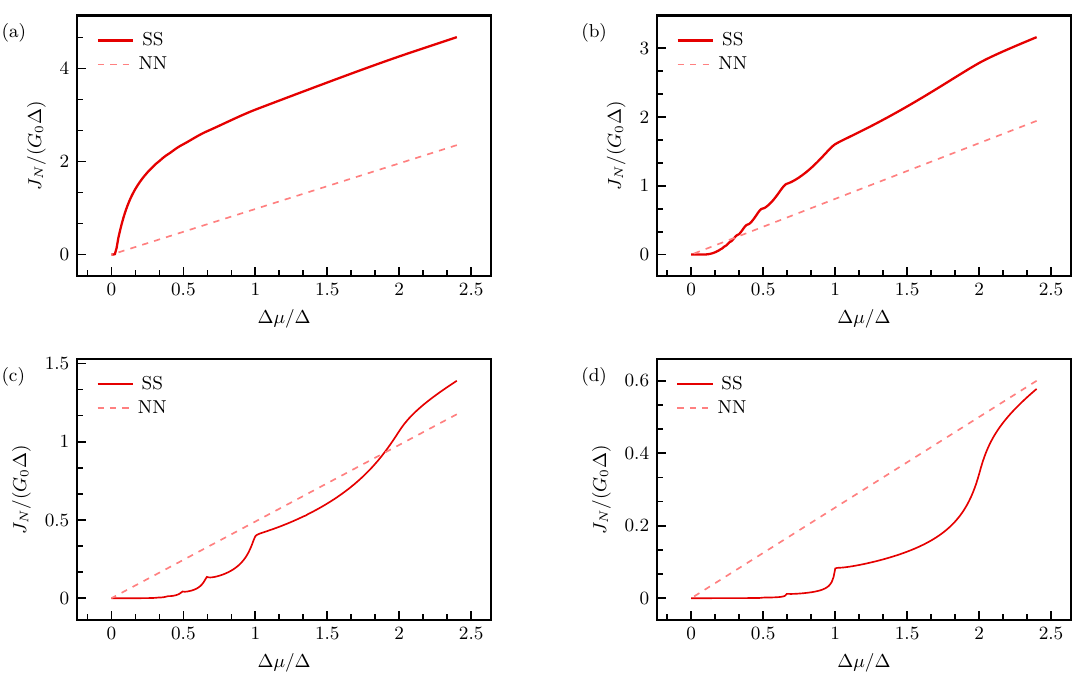}
%\input{plots/Particle}
%\subfigure[]{
%  \includegraphics[width=0.32\textwidth]{image/09_Particle_SS.png}
%  \label{fig:particle_a}
%}\hfill
%\subfigure[]{
%  \includegraphics[width=0.32\textwidth]{image/07_Particle_SS.png}
%  \label{fig:particle_b}
%}\hfill
%\subfigure[]{
%  \includegraphics[width=0.32\textwidth]{image/05_Particle_SS.png}
%  \label{fig:particle_c}
%}
\caption{
Particle current as a function of chemical potential bias for intermediate junction transparencies. 
Solid lines correspond to SS junctions, while dashed lines denote NN junctions. 
From left to right, the panels show:
(a) $\mathcal{T}=0.99$;
(b) $\mathcal{T}=0.9$;
(c) $\mathcal{T}=0.7$;
(d) $\mathcal{T}=0.5$. At lower transparency, the activation of successive multiple Andreev reflection (MAR) processes becomes clearly resolved. 
These appear as distinct peaks in the current--voltage characteristics as the chemical potential gradient is increased, reflecting the opening of additional MAR transport channels by going towards higher $\Delta \mu$.}
\label{fig:Particle_low}
\end{figure*}

\begin{figure*}[t]
\centering
%\hspace{-1cm}
%\input{plots/Zoom_Entropy}
\hspace{-1cm}\includegraphics[width=1.00\textwidth]{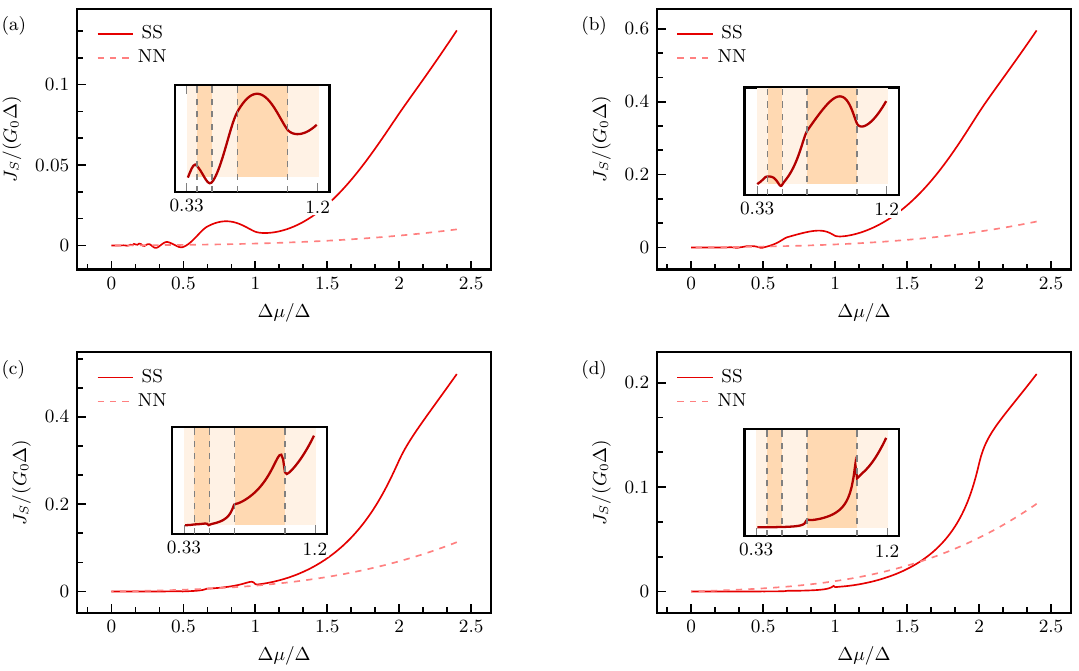}
%\subfigure[]{
%  \includegraphics[width=0.48\textwidth]{image/099_Entropy_SS_Zoom.png}
%  \label{fig:entropy_a}
%}\hfill
%\subfigure[]{
%  \includegraphics[width=0.48\textwidth]{image/09_Entropy_SS_Zoom.png}
%  \label{fig:entropy_b}
%}
%
%\vspace{0.3cm}
%
%\subfigure[]{
%  \includegraphics[width=0.48\textwidth]{image/07_Entropy_SS_Zoom.png}
%  \label{fig:entropy_c}
%}\hfill
%\subfigure[]{
%  \includegraphics[width=0.48\textwidth]{image/05_Entropy_SS_Zoom.png}
%  \label{fig:entropy_d}
%}
\caption{Entropy current as a function of bias for different values of the junction transparency~$\mathcal{T}$.
Solid lines correspond to SS junctions, whereas dashed lines denote NN junctions. 
The insets provide a magnified view of the voltage range where the first oscillatory features emerge upon decreasing the bias.
Vertical dashed lines indicate the onset of multiple Andreev reflection (MAR) processes at 
$\Delta\mu = 2\Delta/n$, with $n \in \mathbb{N}$. 
For clarity, we explicitly mark the thresholds corresponding to $n \in \{2,3,4,5\}$. 
Dark-shaded regions highlight bias intervals where even MAR (eMARs) processes, i.e., \emph{Andreev composite tunneling}, dominate (in the low transparency regime), 
while light-shaded regions indicate dominance of odd MAR (oMARs) processes, i.e., \emph{Andreev composite reflection}.
Panels correspond to different junction transparencies:
(a) ballistic limit, $\mathcal{T}=0.99$;
(b) $\mathcal{T}=0.9$;
(c) $\mathcal{T}=0.7$;
(d) $\mathcal{T}=0.5$.
}
\label{fig:Entropy_low}
\end{figure*}

To better understand the origin of the oscillatory behavior in the entropy current, it is useful to sharpen the analysis of the transition from low transparency to the ballistic limit briefly mentioned above.
Our analysis of the reduced-transparency case reveals a weakening of the MARs' effects, and as transparency approaches zero, entropy and particle currents are effectively suppressed within the gap region. As shown in Fig.~\ref{fig:Particle_low}, the particle current increases progressively within the superconducting gap as $\Delta\mu$ grows, in agreement with previous results~\cite{Bolech_2005}. This current enhancement is particularly evident at intermediate transparency, where distinct peaks associated with the activation of a new MAR are clearly visible at $\Delta \mu\!=\!2\Delta/n$, by going towards higher chemical potential gradient, with $n\!-\!1$ the number of MARs. Moreover, the particle current also increases with higher transparency. 

On the other hand, the entropy current, shown in Fig.~\ref{fig:Entropy_low}, does not increase monotonically with increasing transparency. This behavior is already apparent in the metallic case. At low transparency, the transmission can be treated perturbatively and at first order depends explicitly on the LDoS, \emph{i.e.} the numerator of Eq.~\eqref{eq:metallic_transmission}. In this regime, the transmission function $\mathcal{T}(\omega)$ directly reflects frequency-dependent features of the band structure, such as the characteristic $\sqrt{\omega + E_F}$ behavior of a quadratic dispersion. As the tunnel-overlap strength $\tau$ increases, higher-order contributions --\,representing multiple scattering events at the contact\,-- become relevant, \emph{i.e.} the denominator of Eq.~\eqref{eq:metallic_transmission} decreases. These processes drive the transmission towards the unitary limit, $\mathcal{T}(\omega) \to 1$, and reduce the influence of the LDoS, as higher order diagrams compensate for the spectral weight.
In the high-transparency limit, $\mathcal{T}(\omega)$ becomes increasingly insensitive to the LDoS, yielding a flatter and more symmetric profile around the mean chemical potential, $\mu = (\mu_c + \mu_d)/2 = 0$. Since the entropy current, $J_S$, is governed by the energy-weighted asymmetry of the transmission, the increased symmetry of $\mathcal{T}(\omega)$ leads to a reduction in its magnitude.

Let us now focus on the most salient result, namely the oscillatory behavior observed deep within the superconducting gap by going towards the ballistic regime. 
At first glance, these oscillations can be attributed to the divergence of the superconducting density of states and to the resulting formation of a gapped, nonmonotonic local density of states (LDoS), in contrast to the square-root behavior characteristic of a normal metal. 
However, as discussed in Appendix~\ref{Truncated_system}, this effect is not the primary source of the oscillations; instead, it merely induces a shift of the oscillatory pattern. 
In the following, we identify and explain the underlying mechanisms responsible for the emergence of the oscillatory behavior. 

To provide a clear interpretation of the entropy current, it is essential to distinguish between two different types of transport processes that contribute to it. In the following, we discuss these processes in order of increasing $\Delta \mu$. Notably, one can clearly differentiate between \emph{Andreev composite reflection}, in which a particle is reflected as a hole within the same reservoir through an odd number of MARs (oMARs), as in Fig.~\ref{fig:Andreev_process_reflection},  and \emph{Andreev composite tunneling}, in which a particle tunnels into the opposite reservoir through an even number of MARs (eMARs), as in Fig.~\ref{fig:Andreev_process}.
The onset of oMAR processes occurs at
\begin{equation}
\label{eq:even}
\Delta\mu = \frac{2\Delta}{2n}, \qquad n \in \mathbb{N}, \; n \geq 1,
\end{equation}
whereas eMAR processes are activated at
\begin{equation}
\label{eq:odd}
\Delta\mu = \frac{2\Delta}{2n+1}, \qquad n \in \mathbb{N}, \; n \geq 1.
\end{equation}
The contribution of a given MAR process is largely determined by the off-diagonal component of the Green function. Once a process is activated, it can undergo multiple Andreev reflections, generating virtual Cooper pairs within the superconducting gap. We have verified that these processes provide the dominant contribution in the ballistic channel limit, in agreement with Ref.~\cite{Cuevas_1996}. As the chemical potential difference increases, the corresponding energy window for a given MAR process broadens, leading to an enhancement of its contribution.
However, as $\Delta\mu$ increases further and additional MAR channels become energetically accessible, the contribution of a specific MAR process is progressively reduced. This reduction originates from the increasing involvement of virtual Cooper pairs associated with states outside the superconducting gap, (together with a diminished contribution from those within the gap \footnote{When transport mediated by Cooper pairs created inside the gap is fully suppressed, one can identify a soft upper bound for the eMAR and oMAR processes. After this point, their contribution to the current is reduced, especially in the ballistic limit. These bounds are given by $
\Delta \mu = \frac{2\Delta}{2n+2}
\quad \text{and} \quad
\Delta \mu = \frac{2\Delta}{2n+3},$ for eMAR and oMAR processes, respectively. 
In particular, for a $2n$-eMAR ($(2n+1)$-oMAR) process this corresponds to the activation of the $(2n+2)$-eMAR ($(2n+3)$-oMAR) process.}), whose contribution is strongly suppressed by the pairing correlations of the system, particularly in the ballistic regime \footnote{A particularly relevant case is that of a single Andreev reflection. In this regime, the transport window for Cooper pairs within the superconducting gap is not reduced, and persists even at $\Delta \mu \gg \Delta$.}. These results are consistent with Fig.~2 of Ref.~\cite{Cuevas_2003}.

\begin{figure}[ht]
    \centering
    \includegraphics[width=0.49\textwidth]{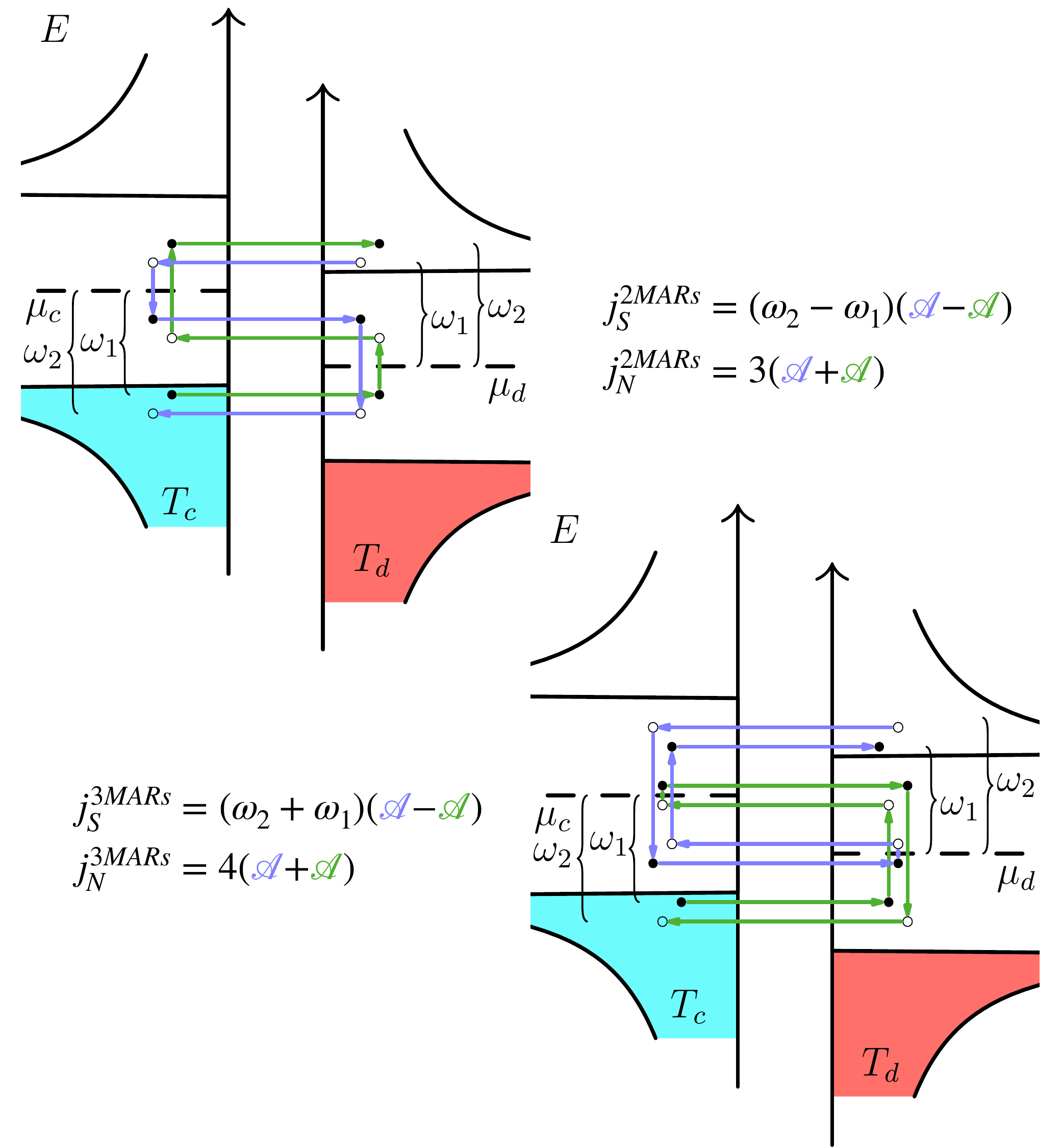} % Replace with your image file name
    \caption{Description of the contributions to entropy and particle current for the \emph{Andreev composite tunneling} (top panel, eMARs) and for the \emph{Andreev composite reflection} (bottom panel, oMARs). On the side of each illustration, the contributions from two isolated processes are shown for both entropy and particle current (denoted as $j_{S/N}^{n_{pair}MARs}$). These two processes are chosen such that their contributions to the entropy current cancel in the case of linear dispersion. The curly symbol $\mathscr{A}$ highlights the contribution of the selected process: the light blue one indicates the transport of a hole from the right reservoir, while the light green corresponds to particle transport originating from the left reservoir. For linear dispersion, these two contributions are equal and therefore cancel in the entropy current. Notice that the processes shown in the top panel occur only if $\Delta\mu > 2\Delta/3$ , while the processes in the bottom panel for $\Delta \mu > \Delta/2$. Additionally, since (for any given $\Delta\mu$) $\omega_2$ is determined by $\omega_1$, the frequency integration (performed over $\omega_1$) must be restricted to $\omega_1 < \omega_2$ in order to avoid double-counting when computing the currents (if the processes are grouped in pairs).}
    \label{fig:Andreev_process_reflection_conclusions}
\end{figure}

The importance of distinguishing between these two types of processes originates on their different contributions to the entropy integrand in Eq.~\eqref{eq:current_total}. In Fig.~\ref{fig:Andreev_process_reflection_conclusions}, we show one example of an eMAR process (upper panel) and another of an oMAR process (lower panel). 
For each type of process, we consider the contribution from two specific frequency channels, such that in the case of a linear dispersion, their sum vanishes. Indeed, the light-blue and light-green transport contributions, $\mathscr{A}$, depend on the densities of states of the particles involved. In the case of linear dispersion, those are equal and, therefore, yield a vanishing contribution to the entropy current ($j^{n_{\mathrm{pair}}\mathrm{MARs}}_{S}$). 
In contrast, for the particle current ($j^{n_{\mathrm{pair}}\mathrm{MARs}}_{N}$), these contributions add up, resulting in a finite value.

Indeed, beyond the part arising from transport channels, $\mathscr{A}$, the entropy current also includes a contribution associated with heat flow.
On the one hand, the particle current is characterized by a factor of $n=3$ and $n=4$, reflecting the number of particles transferred in these two examples. 
By contrast, for entropy transport one must account for the heat carried by each process.
For eMARs, the heat associated with the outgoing particle from the left reservoir and that of the incoming particle from the right reservoir are subtracted. In the case of oMARs, instead, the heat carried by the outgoing particle and the reflected hole add up. As a consequence, oMARs processes provide a larger heat contribution.

We can now understand the oscillatory behavior of the entropy current by focusing on the ballistic limit. As discussed above, MAR processes involving a smaller number of Cooper pairs possess wider transport windows in frequency space. If even and odd MAR processes contributed with the same integral kernel, as for the particle current, the resulting behavior would be monotonic and dictated solely by the progressive enlargement of the transport frequency windows as $\Delta\mu$ increases.
However, for the entropy current the oMAR processes provide a larger heat contribution than the eMAR ones. Consequently, when the oMAR contribution is reduced, the total entropy current decreases until the next oMAR process is activated. This interplay between the relative weights of oMAR and eMAR processes leads to the oscillatory behavior observed in the entropy current.

A special case is the \emph{single Andreev reflection}, shown in Fig.~\ref{fig:Andreev_process_reflection}. This process can be isolated in the case of an NS junction, as illustrated in Fig.~\ref{fig:Particle_Entropy_current_SN} (right panel), and shown to produce a quadratic-like behavior of the entropy current. In the case of particle transport, this mechanism gives rise to the well-known \textit{excess current} \cite{Visuri_2023, klapwijk1982}.
For the entropy current, the situation is different. This process produces a contribution that grows monotonically with increasing bias. The reason is that, in contrast to the particle current, the heat contribution increases as the energy moves away from the chemical potential of the reservoir. This explains the large increasing entropy current contribution observed at high bias for both SN (see Fig.~\ref{fig:Particle_Entropy_current_SN}) and SS junctions (see Fig.~\ref{fig:Entropy_low}).

\begin{figure}[t]
    \centering
    %\hspace{-1cm}
    %\input{plots/099_SS_TotalEntropy}
    \hspace{-1cm}\includegraphics[width=0.5\textwidth]{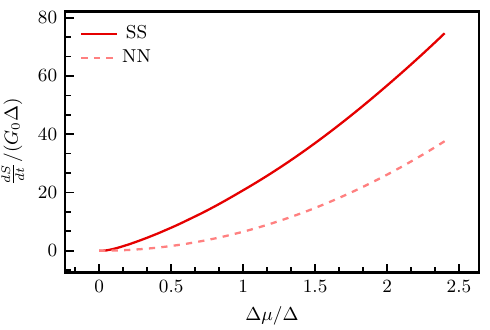}
    % Replace with your image file name
    \caption{\textcolor{black}{The total entropy production for the ballistic case limit of $\mathcal{T} = 0.99$, both for superconducting and normal junctions. The above plot was obtained by directly computing the total entropy production as the sum of the entropy variations in the two reservoirs. This result is consistent with that obtained from the thermodynamic relation~\eqref{eq:Total_entropy_text}}.} 
\label{fig:Total_entropy}
\end{figure}

In summary, the competition between the two mechanisms of \emph{Andreev composite reflection} and \emph{Andreev 
composite tunneling} (with odd or even number of MARs, respectively), is responsible for the predicted rich nonmonotonic behavior of the entropy current.
\textcolor{black}{
\section{Entropy production}
For completeness, we also compute the total entropy production of the entire system. Starting from Eq.~\eqref{eq:total_entropy}, and considering the case in which both reservoirs are at the same temperature, we obtain
\begin{equation}
\label{eq:Total_entropy_text}
    \frac{dS}{dt} = \frac{\Delta \mu}{T} J_N,
\end{equation}
which is always positive, since the chemical potential difference has the same sign as the particle current. Indeed, particles flow from the reservoir with higher chemical potential to the one with lower chemical potential. This is in agreement with the second law of thermodynamics.
In our particular scenario, the distinction between oMAR and eMAR processes done to explain the entropy current, does not apply to the case of the total entropy production. Indeed, in the latter quantity, the two composite processes contribute equally. While in the calculation of the entropy current the two frequency contributions considered in Fig.~\ref{fig:Andreev_process_reflection_conclusions} are added for the eMAR processes and subtracted for the oMAR processes, in the total entropy production, the two frequencies always add. This can also be understood from the fact that, in the present case, the total entropy production depends only on the particle current, which indeed does not need a distinction between even and odd MARs processes. For completeness, we show the behavior of the total entropy production in the ballistic limit in Fig.~\ref{fig:Total_entropy}, which indeed follow the behavior of \eqref{eq:Total_entropy_text}.
}

\textcolor{black}{Therefore, we see that the system exhibits a finite energy production out of equilibrium. This can be traced back to the voltage-driven breaking of time-reversal symmetry. Indeed, under time reversal, the Hamiltonian remains invariant,
\[
\mathbb{T} H \mathbb{T}^{-1} = H,
\]
whereas the matter and energy currents change sign, which implies a change of sign also for the entropy current,
\[
\left.
\begin{aligned}
\mathbb{T} J_N \mathbb{T}^{-1} &= -J_N\\
\mathbb{T} J_E \mathbb{T}^{-1} &= -J_E
\end{aligned}
\right\}
\quad \Longrightarrow \quad
\mathbb{T} J_S \mathbb{T}^{-1} = -J_S.
\]
As a result, the non-equilibrium steady state is mapped onto a different state with opposite currents, while the Hamiltonian is left unchanged. This breaking of time-reversal symmetry gives rise to a finite entropy production in the system \cite{PhysRevE.96.052124, https://doi.org/10.1002/andp.202000276, RevModPhys.93.035008}. The latter indicates that the system is in a non-equilibrium state and breaks the symmetry by selecting a preferred direction for the matter and entropy currents.}

\section{Conclusions and outlook}

Focusing on the behavior of the DC entropy current within the superconducting gap, we find pronounced oscillations arising from MARs, revealing a rich structure of subgap entropy transport that goes well beyond a simple monotonic response. 

From an experimental perspective, recent advances in the direct measurement of entropy in mesoscopic systems—via Maxwell relations~\cite{PhysRevLett.129.227702,Hartman2018} and the Seebeck coefficient~\cite{doi:10.1021/acs.nanolett.1c03591} (i.e., thermopower)—have enabled direct access to the entropy and heat of a variety of systems.
In a cold-atoms context, a comparison with recent experiments \cite{Fabritius_2024, mohan2024universalentropytransportfar} shows that, while the mean-field BCS approach successfully captures the particle current, the entropy current exhibits clear quantitative deviations. 
This highlights the enhanced sensitivity of entropy transport to correlation effects beyond mean field.

A key difference between our BCS-based model and the phenomenological approach reported in \cite{mohan2024universalentropytransportfar}, which describes an ultracold Fermi gas at unitary, lies in the nature of the driving forces for particle transport. In the metallic scenario, the particle current is driven by a combination of chemical potential and temperature gradients, expressed as $\Delta \mu + \alpha_c \Delta T$ through the Seebeck coefficient $\alpha_c$. The latter can be interpreted as the average entropy transported per particle. In contrast, our results indicate that such a relation does not hold in the BCS regime. Specifically, $\Delta \mu$ and $\Delta T$ influence the model junction in fundamentally different ways. Within the Keldysh action framework (see Sec.~\ref{Keldysh Green functions}), one may introduce a temperature gradient between the left and right reservoirs; however, in the absence of a chemical potential bias, this alone is insufficient to induce multiple Andreev reflections (MARs).  This points at a distinct mechanism for transport in the BCS side. We note that our results may also be of significant interest to the condensed matter community. For instance, the SN junctions can be experimentally realized in scanning tunneling microscopy (STM) setups \cite{chen2021introduction,Levi1997}.

Furthermore, our BCS model does not exhibit a proportionality between entropy and particle currents, which was a key feature of the unitary Fermi system analyzed in Ref.~\cite{mohan2024universalentropytransportfar}. Even though the particle current in our BCS model accurately reproduces the experimental behavior with respect to the chemical potential bias, its dependence on the temperature gradient, as well as the oscillating behavior of the entropy current, differ significantly from the observations in the unitary case.
We emphasize that the cited experiments are performed in cold-atom setups in the unitary Fermi gas regime, which is intrinsically distinct from the weak-coupling BCS framework employed in our analysis; this fundamental difference naturally limits a direct quantitative comparison, while at the same time providing a clear direction for future theoretical extensions.
Consistently, in the ballistic limit (see the right panel of Fig.~\ref{fig:Particle_Entropy_current_superconductor}) the entropy current is found to be approximately two orders of magnitude smaller than the experimentally observed values \cite{mohan2024universalentropytransportfar}, suggesting that additional contributions --\,such as pairing fluctuations or beyond–mean-field effects\,-- play an essential role.

In this work, we go beyond the linear dispersion approximation, which inherently imposes particle--hole symmetry and therefore inevitably suppresses the heat current. To overcome this limitation, we consider the full quadratic dispersion relation.
Within the mean-field regime, we further demonstrate that the subgap oscillations of the entropy current can be attributed to two distinct microscopic mechanisms characterized by different heat contributions: \emph{Andreev composite reflection}, in which a particle is effectively reflected as a hole with opposite spin within the same reservoir, and \emph{Andreev composite tunneling}, in which a particle ultimately tunnels into the opposite reservoir.
The former process corresponds to an even number of ``cotunneling particles'' (or, equivalently, an odd number of MARs) where the heat of the particle and the reflected hole contribute additively.
In contrast, \textit{Andreev composite tunneling}, associated with an odd number of ``cotunneling particles'' (or an even number of MARs), is determined by the difference between the heat of the outgoing and incoming particles in the two reservoirs.
 Both of these processes combined create the nontrivial behavior that we compute. 

Considering deviations from linear dispersion is not the only route to avoid the suppression of the heat current due to particle-hole symmetry. One could alternatively consider two different superconducting gaps, $\Delta_c \neq \Delta_d$, with either linear or quadratic dispersion. A brief calculation shows that this choice can produce, if $|\Delta_c - \Delta_d|$ is sufficiently large, a distinct and larger contribution as compared to the oscillations observed in the quadratic-dispersion case which was our focus here. This suggests that further investigations along this direction may be worthwhile.

\section*{ACKNOWLEDGMENTS}

We thank T. Esslinger group, S. Uchino, A.M. Visuri, S. Sur, P. Kattel and C. Berthod for fruitful discussion. This work was supported by the Swiss National Science Foundation under Division II (Grant No. 200020-219400).

\section*{DATA AVAILABILITY}

The data that support the findings of this article are openly available \cite{bertolusso_2026_22014928}.

\appendix

\section{Thermodynamics}
\label{Thermodynamics}

In this appendix we briefly derive the thermodynamic transport equations from which we can extract the particle and entropy currents. In the following, we will identify the particle $c$ with the left reservoir ($L$) and the particle $d$ with the right reservoir ($R$). We start by defining the particle and entropy currents as \cite{mahan_many-particle_2000, Mohan2024Universal},
\begin{equation}
\begin{aligned}
    J_N &= -\frac{1}{2}\frac{d\Delta N}{dt} = -\frac{1}{2} \left(\frac{dN_L}{dt} - \frac{dN_R}{dt}  \right) \\
    J_S &= -\frac{1}{2}\frac{d\Delta S}{dt} = -\frac{1}{2} \left(\frac{dS_L}{dt} - \frac{dS_R}{dt}  \right)
\end{aligned}
\end{equation}
We shall also introduce two conserved quantities, the total number of atoms and the total internal energy, respectively,
\begin{equation}
\begin{aligned}
    \frac{dN_L}{dt} + \frac{dN_R}{dt} &= 0
    \\
    \frac{dU_L}{dt} + \frac{dU_R}{dt} &= 0
\end{aligned}
\end{equation}
\textcolor{black}{By taking the mean value of the above relations, we are going to prove them for our particular quantum system in the following. If we compute the particle flow from the left reservoir, we get,
\begin{equation}
     J^{(c)}_N 
    = -i \left\langle \sum_{\textbf{k},\textbf{k}',\sigma} 
    \tau \Big[
        e^{i\Delta \mu t'} 
        \bar{\psi}_{c\sigma,\textbf{k}}\,
        \psi_{d\sigma,\textbf{k}'}
        - \text{h.c.}
    \Big] \right\rangle
\end{equation}
as well as the one for the right,
\begin{equation}
     J^{(d)}_N 
    = -i \left\langle \sum_{\textbf{k},\textbf{k}',\sigma} 
    \tau \Big[
        e^{-i\Delta \mu t'} 
        \bar{\psi}_{d\sigma,\textbf{k}}\,
        \psi_{c\sigma,\textbf{k}'}
        - \text{h.c.}
    \Big] \right\rangle
\end{equation}
we notice that,
\begin{equation}
    J^{(c)}_N = - J^{(d)}_N
\end{equation}
which proves the particle conservation.
\begin{widetext}
The energy flow instead is,
\begin{equation}
\begin{aligned}
\label{eq:energy_flow}
     J^{(c/d)}_U 
    &= i \left\langle 
\left[
H,\ 
\sum_{\textbf{k},\sigma} 
\epsilon_{\textbf{k}}\,
\bar{\psi}_{c/d\sigma,\textbf{k}}\, 
\psi_{c/d\sigma,\textbf{k}} + \sum_{\textbf{k}}
             \Delta\,
             \psi_{c/d\uparrow,\mathbf{k}}\,
             \psi_{c/d\downarrow,-\mathbf{k}}
              + \text{h.c.} 
\right]\right\rangle \\
& = -i\tau \langle \sum_{k,k',\sigma} \epsilon_{\textbf{k}} [e^{\pm i\Delta \mu t} \bar{\psi}_{c/d,\sigma,\textbf{k}}  \psi_{d/c,\sigma,\textbf{k}'} - h.c.] + \sum_{k,k'} \Delta [e^{\pm i\Delta \mu t}\psi_{d/c,\uparrow,\textbf{k}'} \psi_{c/d,\downarrow,\textbf{-k}} - e^{\pm i \Delta \mu t} \psi_{d/c,\downarrow,\textbf{-k}'} \psi_{c/d,\uparrow,\textbf{k}}  -h.c.] \rangle
\end{aligned}
\end{equation}
to get the total energy variation, we sum the contributions from the $c$ and $d$ reservoirs,
\begin{equation}
    J^{c}_U + J^{d}_U =  \sum_{k,k'} \Delta e^{ i\Delta \mu t} \left[ \langle \psi_{d,\uparrow,\textbf{k}'} \psi_{c,\downarrow,\textbf{k}} - \psi_{d,\downarrow,\textbf{k}'} \psi_{c,\uparrow,\textbf{k}} \rangle\right] + \Delta e^{-i\Delta \mu t} \left[\langle \psi_{c,\uparrow,\textbf{k}'} \psi_{d,\downarrow,\textbf{k}} - \psi_{c,\downarrow,\textbf{k}'} \psi_{d,\uparrow,\textbf{k}}\rangle\right] - h.c. = 0
\end{equation}
when summing the two energy flows from the $c$ and $d$ reservoirs, the first term in \eqref{eq:energy_flow} cancels out, and since the system is symmetric under spin rotation, the above expression gives zero. For completeness, we numerically verify the conservation of particle and entropy current for the superconducting system with a fixed value of MARs process.
\end{widetext}}
Thus, the corresponding currents can be written referring to a single reservoir. For instance,
\begin{equation}
\label{eq:energy_particle_current}
\begin{aligned}
    J_{U} &= -\frac{1}{2}\frac{d\Delta U}{dt} = -\frac{1}{2} \left(\frac{dU_L}{dt} - \frac{dU_R}{dt}  \right) = \frac{dU_R}{dt} \\
    J_{N} &= -\frac{1}{2}\frac{d\Delta N}{dt} = -\frac{1}{2} \left(\frac{dN_L}{dt} - \frac{dN_R}{dt}  \right) = \frac{dN_R}{dt}
\end{aligned}
\end{equation}
and by using the first law of thermodynamics \cite{Callen:450289} for a generic reservoir $I$,
\begin{equation}
    dQ_I = T_IdS_I = dU_I - \mu_I dN_I
\end{equation}
we find,
\begin{equation}
    \frac{dS_I}{dt} = \frac{1}{T_I}\frac{dU_I}{dt} - \frac{\mu_I}{T_I} \frac{dN_I}{dt}
\end{equation}
From the energy current, we can derive the heat current, (meaning the variation of the grand canonical Hamiltonian of Eq.~\ref{eq:Hamiltonian}),
\begin{equation}
\label{heat}
    J_H = J_U - \mu J_N = - \frac{1}{2}\frac{d\Delta Q}{dt}
\end{equation}
combining this with \eqref{eq:energy_particle_current}  we find \cite{Mohan2024Universal},
\begin{equation}
\begin{aligned}
\label{entropy}
    J_S & = -\frac{1}{2}\left( \frac{dS_L}{dt} - \frac{dS_R}{dt}\right) \\
    & = -\frac{1}{2} \left[\left( -\frac{1}{T_L} J_U + \frac{\mu_L}{T_L} J_N \right) - \left( \frac{1}{T_R} J_U - \frac{\mu_R}{T_R} J_N\right)\right] \\
    & = \frac{1}{4T^2 - \Delta T^2}
    \left(4T\,J_H + \Delta\mu\,\Delta T\,J_N\right)
\end{aligned}
\end{equation}
where
\begin{align*}
    T &= (T_L +T_R)/2,\quad &\mu &= (\mu_L +\mu_R)/2\\
    \Delta T &= T_R - T_L,\quad &\Delta \mu &= \mu_L - \mu_R
\end{align*}

Furthermore, we generalize Eq.~\eqref{eq:energy_particle_current} to quantum thermodynamic quantities, the energy and particle currents are,
\begin{equation}
\begin{aligned}
\label{quantum_thermodynamics}
     J_U  &= -\frac{1}{2}\left(\left\langle\frac{d U_L}{dt}\right\rangle - \left\langle\frac{d U_R}{dt}\right\rangle \right) \\
     J_N  &= -\frac{1}{2}\left(\left\langle\frac{d N_L}{dt}\right\rangle - \left\langle\frac{d N_R}{dt}\right\rangle \right)
\end{aligned}
\end{equation}
where $U_I$ is the Hamiltonian in the canonical ensemble for reservoir $I$ and $N_I$ the number operator. 

Using Eqs.~\eqref{heat},~\eqref{entropy},~\eqref{quantum_thermodynamics}, we derive the expression for the entropy current in the quantum case.

\textcolor{black}{Additionally, we can also compute the total entropy variation \cite{RevModPhys.93.035008},
\begin{equation}
\label{eq:total_entropy}
    \frac{dS}{dt} = \frac{dS_L}{dt} + \frac{dS_R}{dt} = \frac{\Delta T J_S + \Delta \mu J_N}{T}
\end{equation}
}
\newline
\section{Transmission function}
\label{Transperancy}

Here we derive the transmission function in the metallic limit, and then implement the same definition for the superconducting case.
 
In order to define the probability of tunneling from one reservoir to the other, and therefore the physical parameter to tune, we start from the Landauer-Büttiker formula for the particle current \cite{Landauer_1957},
\begin{equation}
\frac{J_N}{G_0\Delta} = \int dE \mathcal{T}^2(E) \mathcal{N}(E) v(E) [n_L(E) - n_R(E)]
\end{equation}
which describes the particle current between two reservoirs, left ($L$) and right ($R$).
Here the integration is performed over the entire energy spectrum. $\mathcal{N}(E)$ denotes the density of states and $v(E)$ the particle velocity, such that $\mathcal{N}(E)\,v(E)=1$. The functions $n_{L/R}$ are the Fermi–Dirac distributions of the left/right reservoirs, respectively. $\mathcal{T}^2(E)$ denotes the tunneling (transmission) probability.
The tunable parameter of the system is the transmission function $\mathcal{T}^2(E)$; for instance, in a ballistic channel one has $\mathcal{T}(E)=1$, if instead the channel is closed, $\mathcal{T}(E)=0$.
Therefore, if we impose the transmission function to be our tunable parameter at the Fermi level $\mathcal{T}(0)$ and allow it to change from low to high transparency, we will need to solve the following equation to find the tunneling coupling,
\begin{equation}
\label{T_normalized}
 \mathcal{T}^2 \equiv \mathcal{T}^2(0) = \frac{4 \tau^2 \pi^2 \rho^2(0)}{(1-\tau^2 G^R(0)^2)(1-\tau^2 G^A(0)^2)} 
\end{equation}
where $\mathcal{T}$ is defined in the metallic limit from Eq.~\eqref{eq:current_formula_metal}.
In the linear dispersion case we have $G^R(0) = -i \pi \rho(0)$. Therefore, the above expression reduces to
\begin{equation}
 \mathcal{T}^2 = \frac{4 (\tau/W)^2)}{(1+(\tau/W)^2)^2} 
\end{equation}
where $W = \frac{1}{\pi \rho(0)}$.
By solving from Eq.~(\ref{T_normalized}) for a general dispersion, we obtain the tunneling coupling
\begin{widetext}
\begin{equation}
\tau = \sqrt{\frac{\mathcal{T}^2}{\mathcal{T}^2 x+2 \Im{G^R}^2+\sqrt{-\mathcal{T}^4 y^2+4 \mathcal{T}^2 x \Im{G^R}^2+4 \Im{G^R}^4}}}
\end{equation}
\end{widetext}
where
$$
G^{R/A} \equiv G^{R/A}(0)
$$
$$
x = \Re{G^R}^2 - \Im{G^R}^2
$$
$$
y = 2\Re{G^R}\Im{G^R} 
$$
Note that the retarded and advanced Green functions have identical real and opposite imaginary parts.

\begin{figure}
    \centering
    \includegraphics[width=0.3\textwidth]{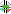} % Replace with your image file name
    \caption{\textcolor{black}{ Schematic representation of a possible fifth-order oMAR process, where we delineate the frequency structure as described by Eq.~\eqref{eq:Omega_tower}. Starting from a spin-up particle at frequency $\omega_0$ (say, between the two chemical potentials), a cascade of frequencies is generated as a result of multiple Andreev reflections. The inverse Green function in Eq.~\eqref{eq:Super_green_4order} can describe this process up to the fourth order (solid lines), but to complete the oMAR transport process one would need to truncate at a higher order (dashed lines).}}
    \label{fig:Omega_tower}
\end{figure}

{\color{black}
\section{The superconducting action}
\label{Action_structure}
Following Ref.\,\onlinecite{Visuri_2023}, we construct the action for the all superconducting point contact system. As we already outlined in the main text, the action is not block diagonal in frequency and this is physically translated as MARs processes. In this Appendix, we explicitly show the latter by constructing the inverse Green function in the frequency space. First of all we need to construct the frequency structure of the action. The action for the point contact reads as follows,
\begin{equation}
    S = \int d\omega \bar{\mathbf{\Psi}}(\omega) \mathcal{G}^{-1}(\omega)\mathbf{\Psi}(\omega)
\end{equation} 
Where the field, in the Keldysh formalism, is,
\begin{widetext}
\begin{equation}
    \bar{\mathbf{\Psi}}(\omega) = \left( \bar{\psi}^{cl}_{c,\uparrow}(\bar{\omega}_0), \bar{\psi}^{q}_{c,\uparrow}(\bar{\omega}_0), \bar{\psi}^{cl}_{d,\uparrow}(\bar{\omega}_0), \bar{\psi}^{q}_{d,\uparrow}(\bar{\omega}_0), \psi^{cl}_{c,\downarrow}(\bar{\omega}_1), \psi^{q}_{c,\downarrow}(\bar{\omega}_1), ... \right)
\end{equation}
and the inverse Green function truncated at the fourth MARs process is,
\newline
\begin{equation}
\label{eq:Super_green_4order}
\scalebox{0.8}{$
\mathcal{G}^{-1}(\bar{\omega}_0) =
\left[
\begin{array}{@{}*{10}{c}@{}}
\Omega_{c,\uparrow}(\bar{\omega}_0) & \mathrm{T} & \Delta_{c}(\bar{\omega}_0) & 0 & 0 & 0 & 0 & 0 & 0 & 0 \\
\mathrm{T} & \Omega_{d,\uparrow}(\bar{\omega}_0) & 0 & 0 & \Delta_{d}(\bar{\omega}_0) & 0 & 0 & 0 & 0 & 0 \\
\Delta_{c}(\bar{\omega}_0) & 0 & \Omega_{c,\downarrow}(\bar{\omega}_1) & -\mathrm{T} & 0 & 0 & 0 & 0 & 0 & 0 \\
0 & 0 & -\mathrm{T} & \Omega_{d,\downarrow}(\bar{\omega}_1) & 0 & 0 & \Delta_{d}(\bar{\omega}_2) & 0 & 0 & 0 \\
0 & \Delta_{d}(\bar{\omega}_0) & 0 & 0 & \Omega_{d,\downarrow}(\bar{\omega}_{-1}) & -\mathrm{T} & 0 & 0 & 0 & 0 \\
0 & 0 & 0 & 0 & -\mathrm{T} & \Omega_{c,\downarrow}(\bar{\omega}_{-1}) & 0 & 0 & \Delta_{c}(\bar{\omega}_{-2}) & 0 \\
0 & 0 & 0 & \Delta_{d}(\bar{\omega}_2) & 0 & 0 & \Omega_{d,\uparrow}(\bar{\omega}_2) & \mathrm{T} & 0 & 0 \\
0 & 0 & 0 & 0 & 0 & 0 & \mathrm{T} & \Omega_{c,\uparrow}(\bar{\omega}_2) & 0 & 0 \\
0 & 0 & 0 & 0 & 0 & \Delta_{c}(\bar{\omega}_{-2}) & 0 & 0 & \Omega_{c,\uparrow}(\bar{\omega}_{-2}) & \mathrm{T} \\
0 & 0 & 0 & 0 & 0 & 0 & 0 & 0 & \mathrm{T} & \Omega_{d,\uparrow}(\bar{\omega}_{-2})
\end{array}
\right]
$}
\end{equation}
\end{widetext}
where we define the same-spin block of the inverse Green function as,
\begin{equation}
\begin{aligned}
\Omega_{i \uparrow}(\bar{\omega}) & =\left(\begin{array}{cc}
0 & {\left[G_{i,A}(\bar{\omega})\right]_{\uparrow,\uparrow}^{-1}} \\
{\left[G_{i,R}(\bar{\omega})\right]_{\uparrow,\uparrow}^{-1}} & {\left[G_{i,K}(\bar{\omega})\right]_{\uparrow,\uparrow}^{-1}}
\end{array}\right) \\
& =\left(\begin{array}{cc}
0 & {\left[G_{i,A}(\bar{\omega})\right]_{\downarrow,\downarrow}^{-1}} \\
{\left[G_{i,R}(\bar{\omega})\right]_{\downarrow,\downarrow}^{-1}} & {\left[G_{i,K}(\bar{\omega})\right]_{\downarrow,\downarrow}^{-1}}
\end{array}\right)=\Omega_{i \downarrow}(-\bar{\omega}),
\end{aligned}\\
\end{equation}
\newline
The latter constructs the main diagonal block, and we identify it as the tunneling block, \textit{i.e.} the horizontal line in Fig.~\ref{fig:Andreev_process}, where a particle tunnels by being annihilated and creating another particle with the same spin on the opposite reservoir, with the tunneling matrix,
\newline
\begin{equation}
    \begin{aligned}
&\begin{aligned}
\mathrm{T} & =\left(\begin{array}{cc}
0 & -\tau \\
-\tau & 0
\end{array}\right) \\
\end{aligned}
\end{aligned}
\end{equation}
\newline
The outer off-diagonal blocks are built by the anomalous component due to the superfluid pairing, \textit{i.e.} the vertical lines in Fig.~\ref{fig:Andreev_process}, and read as
\begin{equation}
\begin{aligned}
\Delta_i(\bar{\omega}) & =\left(\begin{array}{cc}
0 & {\left[G_{i,A}(\bar{\omega})\right]_{\uparrow,\downarrow}^{-1}} \\
{\left[G_{i,R}(\bar{\omega})\right]_{\uparrow,\downarrow}^{-1}} & {\left[G_{i,K}(\bar{\omega})\right]_{\uparrow,\downarrow}^{-1}}
\end{array}\right)
\end{aligned}
\end{equation}
\newline
The sequence of frequencies in Eq.\,(\ref{eq:Super_green_4order}) is generated according to the following recursive relations \cite{Bolech_2005,Visuri_2023},
\begin{equation}
\label{eq:Omega_tower}
\begin{aligned}
\omega_n &= 2\mu_{2-n \bmod 2}-\omega_{n-1}, \\
\omega_{-n} &= 2\mu_{1+n \bmod 2}-\omega_{1-n}.
\end{aligned}
\end{equation}
These frequencies are subsequently shifted according to $\bar{\omega}_n=\omega_n-\mu_{c/d}$, depending on whether the field is associated with the left ($c$) or right ($d$) reservoir; cf.~Fig.~\ref{fig:Andreev_process_reflection_conclusions}. Notice how these shifted frequencies grow continuously, eventually helping to motivate the matrix truncation \cite{Bolech_2005}. All the examples of processes given in the main-text figures are fully captured at the level of truncation shown in Eq.~(\ref{eq:Super_green_4order}).
As another example, Fig.\,\ref{fig:Omega_tower} shows a full frequency sequence generated up to the fourth order (and extended to a fifth in dashed lines). 

Once the inverse Green function has been constructed, we obtain the corresponding Green function by matrix inversion and extract the integrand of Eq.~\eqref{eq:current_final_super}. The resulting integral is then evaluated numerically using a quadrature method that adaptively samples $\omega_0$ \cite{1983qspa.book.....P}.}
{\color{black}
\section{Currents}
\label{Currents}
Throughout this Appendix, we briefly summarize the main steps involved in the calculation of the current commutator. Starting from the particle current, we write
\begin{equation}
J^{(c)}_N = i\langle \left[ H_T, N_c \right]\rangle + i\langle \left[ H_c, N_c \right]\rangle
\end{equation}
where $N_c = \sum_{\textbf{k}, \sigma}
\bar{\psi}_{c,\sigma,\textbf{k}},
\psi_{c,\sigma,\textbf{k}}$. The first contribution is
\begin{equation}
i \langle \left[ H_T, N_c \right]\rangle = i \left\langle \sum_{\textbf{k},\textbf{k}',\sigma}
\tau \Big[
e^{i\Delta \mu t'}
\bar{\psi}_{c,\sigma,\textbf{k}},
\psi_{d,\sigma,\textbf{k}'}
- \text{h.c.} \Big] \right\rangle.
\end{equation}
The second term reads
\begin{equation}
\begin{aligned}
&i\langle \left[ H_c, N_c \right]\rangle = i\left[  \sum_{\textbf{k}}
\Delta \langle
\psi_{c,\uparrow,\mathbf{k}},
\psi_{c,\downarrow,-\mathbf{k}} \rangle
+ \text{h.c.}, N_c \right] \\
&= i\sum_{\textbf{k}} \left[\Delta \langle   \psi_{c,\uparrow,\mathbf{k}} \psi_{c,\downarrow,\mathbf{-k}}\rangle - \Delta \langle   \bar{\psi}_{c,\downarrow,\mathbf{-k}} \bar{\psi}_{c,\uparrow,\mathbf{k}}\rangle\right] = 0.
\end{aligned}
\end{equation}
This result follows from our choice of a real-valued superfluid gap. Indeed, $\Delta = g \langle   \bar{\psi}_{c/d,\uparrow,\mathbf{k}} \bar{\psi}_{c/d,\downarrow,\mathbf{-k}}\rangle =  g \langle   \psi_{c/d,\downarrow,\mathbf{k}} \psi_{c/d\uparrow,\mathbf{-k}}\rangle$, where $g$ denotes the contact interaction strength.

For the heat current, we evaluate
\begin{equation}
J^{(c)}_H = -i\langle \left[ H_T, H_c \right]\rangle.
\end{equation}
Using
\begin{widetext}
\begin{equation}
\begin{aligned}
\Dot{\bar{\psi}}_{c\uparrow,\textbf{k}} = i\left[H, \bar{\psi}_{c\uparrow,\textbf{k}}\right]
= i\left[ (\epsilon_{c,\textbf{k}} - \mu_c) \bar{\psi}_{c,\uparrow,\textbf{k}} + \Delta \psi_{c,\downarrow,\textbf{-k}} - \sum_{\textbf{k}'} \tau e^{-i\Delta \mu t} \bar{\psi}_{d\uparrow,\textbf{k}'} \right],
\end{aligned}
\end{equation}
\end{widetext}
we recover the current expressions given in Eqs.~\eqref{eq:particle_current} and \eqref{eq:heat_current} of the main  text.}

\section{Truncated system}
\label{Truncated_system}

\begin{figure}[h!]
    \centering
    %\hspace{-1cm}
    %\input{plots/LDOS_Truncated}
    \hspace{-1cm}\includegraphics[width=0.5\textwidth]{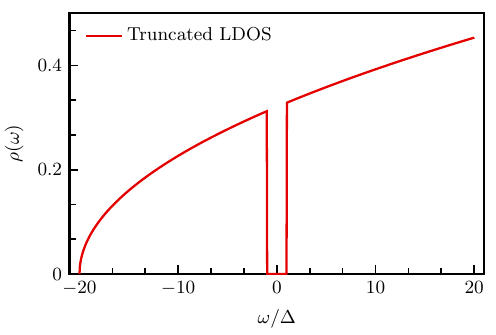}
    \caption{The local density of state (LDoS) for the truncated system at $\omega_{\text{cutoff}}=0$ as a function of the frequency with $E_F = 20\Delta$.}
    \label{fig:LDOS_truncated.png}
\end{figure}

\begin{figure}[]
    \centering
    %\hspace{-1cm}
    %\input{plots/099_Entropy_truncated}
    \hspace{-1cm}\includegraphics[width=0.5\textwidth]{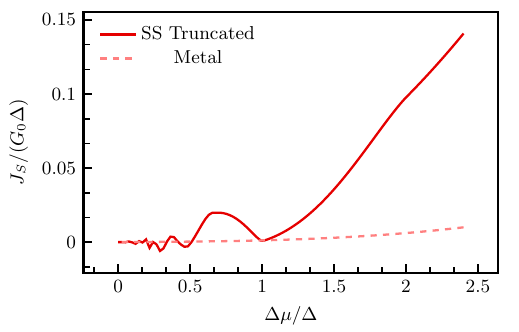}
    \caption{Entropy current  behavior of the truncated SS junction with $\omega_{\text{cutoff}} = 0$, at transparency $\mathcal{T} = 0.99$.}
    \label{fig:099_debye0}
\end{figure}

To elucidate the origin of the entropy behavior, we introduce an artificial model in which Andreev coupling is allowed only within the superconducting gap, while outside the gap the system is assumed to behave as a normal metal.
We therefore introduce a frequency cutoff, denoted by \(\omega_{\mathrm{cutoff}}\), such that the system is effectively metallic or superconducting depending on the energy. Specifically,
\begin{equation}
\begin{cases}
|\omega| > \sqrt{\Delta^2 + \omega_{\mathrm{cutoff}}^2}, & \text{Metal}, \\
|\omega| \leq \sqrt{\Delta^2 + \omega_{\mathrm{cutoff}}^2}, & \text{Superconductor}.
\end{cases}
\end{equation}

This construction allows us to disentangle the origin of the oscillatory behavior and to determine whether it arises from the nonmonotonic divergent LDoS or from the presence of a pairing mechanism confined to the gapped region. We analyze the limit of $\omega_{\text{cutoff}}=0$ in order to avoid the divergence of the LDoS at the gap edges. For completeness we show the LDoS for this system in Fig.~\ref{fig:LDOS_truncated.png}. 

Notably, as we can see from the entropy transport in Fig.~\ref{fig:099_debye0}, 
the oscillatory behavior remains present even in the absence of a divergent LDoS at the gap edge, provided that Cooper-pair formation is restricted to the gapped region, which still permits MAR processes. From the latter, we observed that the physical superconductor density of states only introduced a shift in the peaks of the entropy current in Fig.~\ref{fig:Particle_Entropy_current_superconductor}.
\newline
\newline

\bibliography{apssamp,totphys-A-J,totphys-K-Z}

%apsrev4-2.bst 2019-01-14 (MD) hand-edited version of apsrev4-1.bst
%Control: key (0)
%Control: author (8) initials jnrlst
%Control: editor formatted (1) identically to author
%Control: production of article title (0) allowed
%Control: page (0) single
%Control: year (1) truncated
%Control: production of eprint (0) enabled
\begin{thebibliography}{46}%
\makeatletter
\providecommand \@ifxundefined [1]{%
 \@ifx{#1\undefined}
}%
\providecommand \@ifnum [1]{%
 \ifnum #1\expandafter \@firstoftwo
 \else \expandafter \@secondoftwo
 \fi
}%
\providecommand \@ifx [1]{%
 \ifx #1\expandafter \@firstoftwo
 \else \expandafter \@secondoftwo
 \fi
}%
\providecommand \natexlab [1]{#1}%
\providecommand \enquote  [1]{``#1''}%
\providecommand \bibnamefont  [1]{#1}%
\providecommand \bibfnamefont [1]{#1}%
\providecommand \citenamefont [1]{#1}%
\providecommand \href@noop [0]{\@secondoftwo}%
\providecommand \href [0]{\begingroup \@sanitize@url \@href}%
\providecommand \@href[1]{\@@startlink{#1}\@@href}%
\providecommand \@@href[1]{\endgroup#1\@@endlink}%
\providecommand \@sanitize@url [0]{\catcode `\\12\catcode `\$12\catcode `\&12\catcode `\#12\catcode `\^12\catcode `\_12\catcode `\%12\relax}%
\providecommand \@@startlink[1]{}%
\providecommand \@@endlink[0]{}%
\providecommand \url  [0]{\begingroup\@sanitize@url \@url }%
\providecommand \@url [1]{\endgroup\@href {#1}{\urlprefix }}%
\providecommand \urlprefix  [0]{URL }%
\providecommand \Eprint [0]{\href }%
\providecommand \doibase [0]{https://doi.org/}%
\providecommand \selectlanguage [0]{\@gobble}%
\providecommand \bibinfo  [0]{\@secondoftwo}%
\providecommand \bibfield  [0]{\@secondoftwo}%
\providecommand \translation [1]{[#1]}%
\providecommand \BibitemOpen [0]{}%
\providecommand \bibitemStop [0]{}%
\providecommand \bibitemNoStop [0]{.\EOS\space}%
\providecommand \EOS [0]{\spacefactor3000\relax}%
\providecommand \BibitemShut  [1]{\csname bibitem#1\endcsname}%
\let\auto@bib@innerbib\@empty
%</preamble>
\bibitem [{\citenamefont {Valtolina}\ \emph {et~al.}(2015)\citenamefont {Valtolina}, \citenamefont {Burchianti}, \citenamefont {Amico}, \citenamefont {Neri}, \citenamefont {Xhani}, \citenamefont {Seman}, \citenamefont {Trombettoni}, \citenamefont {Smerzi}, \citenamefont {Zaccanti}, \citenamefont {Inguscio},\ and\ \citenamefont {Roati}}]{Valtolina_2015}%
  \BibitemOpen
  \bibfield  {author} {\bibinfo {author} {\bibfnamefont {G.}~\bibnamefont {Valtolina}}, \bibinfo {author} {\bibfnamefont {A.}~\bibnamefont {Burchianti}}, \bibinfo {author} {\bibfnamefont {A.}~\bibnamefont {Amico}}, \bibinfo {author} {\bibfnamefont {E.}~\bibnamefont {Neri}}, \bibinfo {author} {\bibfnamefont {K.}~\bibnamefont {Xhani}}, \bibinfo {author} {\bibfnamefont {J.~A.}\ \bibnamefont {Seman}}, \bibinfo {author} {\bibfnamefont {A.}~\bibnamefont {Trombettoni}}, \bibinfo {author} {\bibfnamefont {A.}~\bibnamefont {Smerzi}}, \bibinfo {author} {\bibfnamefont {M.}~\bibnamefont {Zaccanti}}, \bibinfo {author} {\bibfnamefont {M.}~\bibnamefont {Inguscio}},\ and\ \bibinfo {author} {\bibfnamefont {G.}~\bibnamefont {Roati}},\ }\bibfield  {title} {\bibinfo {title} {Josephson effect in fermionic superfluids across the {BEC}-{BCS} crossover},\ }\href {https://doi.org/10.1126/science.aac9725} {\bibfield  {journal} {\bibinfo  {journal} {Science}\ }\textbf {\bibinfo {volume} {350}},\ \bibinfo {pages} {1505–1508} (\bibinfo
  {year} {2015})}\BibitemShut {NoStop}%
\bibitem [{\citenamefont {Krinner}\ \emph {et~al.}(2014)\citenamefont {Krinner}, \citenamefont {Stadler}, \citenamefont {Husmann}, \citenamefont {Brantut},\ and\ \citenamefont {Esslinger}}]{Krinner_2014}%
  \BibitemOpen
  \bibfield  {author} {\bibinfo {author} {\bibfnamefont {S.}~\bibnamefont {Krinner}}, \bibinfo {author} {\bibfnamefont {D.}~\bibnamefont {Stadler}}, \bibinfo {author} {\bibfnamefont {D.}~\bibnamefont {Husmann}}, \bibinfo {author} {\bibfnamefont {J.-P.}\ \bibnamefont {Brantut}},\ and\ \bibinfo {author} {\bibfnamefont {T.}~\bibnamefont {Esslinger}},\ }\bibfield  {title} {\bibinfo {title} {Observation of quantized conductance in neutral matter},\ }\href {https://doi.org/10.1038/nature14049} {\bibfield  {journal} {\bibinfo  {journal} {Nature}\ }\textbf {\bibinfo {volume} {517}},\ \bibinfo {pages} {64–67} (\bibinfo {year} {2014})}\BibitemShut {NoStop}%
\bibitem [{\citenamefont {Morpurgo}\ \emph {et~al.}(1998)\citenamefont {Morpurgo}, \citenamefont {Klapwijk},\ and\ \citenamefont {van Wees}}]{Morpurgo}%
  \BibitemOpen
  \bibfield  {author} {\bibinfo {author} {\bibfnamefont {A.~F.}\ \bibnamefont {Morpurgo}}, \bibinfo {author} {\bibfnamefont {T.~M.}\ \bibnamefont {Klapwijk}},\ and\ \bibinfo {author} {\bibfnamefont {B.~J.}\ \bibnamefont {van Wees}},\ }\bibfield  {title} {\bibinfo {title} {Hot electron tunable supercurrent},\ }\href {https://doi.org/10.1063/1.120612} {\bibfield  {journal} {\bibinfo  {journal} {Applied Physics Letters}\ }\textbf {\bibinfo {volume} {72}},\ \bibinfo {pages} {966} (\bibinfo {year} {1998})}\BibitemShut {NoStop}%
\bibitem [{\citenamefont {Del~Pace}\ \emph {et~al.}(2021)\citenamefont {Del~Pace}, \citenamefont {Kwon}, \citenamefont {Zaccanti}, \citenamefont {Roati},\ and\ \citenamefont {Scazza}}]{Del_pace_Tunneling}%
  \BibitemOpen
  \bibfield  {author} {\bibinfo {author} {\bibfnamefont {G.}~\bibnamefont {Del~Pace}}, \bibinfo {author} {\bibfnamefont {W.~J.}\ \bibnamefont {Kwon}}, \bibinfo {author} {\bibfnamefont {M.}~\bibnamefont {Zaccanti}}, \bibinfo {author} {\bibfnamefont {G.}~\bibnamefont {Roati}},\ and\ \bibinfo {author} {\bibfnamefont {F.}~\bibnamefont {Scazza}},\ }\bibfield  {title} {\bibinfo {title} {Tunneling transport of unitary fermions across the superfluid transition},\ }\href {https://doi.org/10.1103/PhysRevLett.126.055301} {\bibfield  {journal} {\bibinfo  {journal} {Phys. Rev. Lett.}\ }\textbf {\bibinfo {volume} {126}},\ \bibinfo {pages} {055301} (\bibinfo {year} {2021})}\BibitemShut {NoStop}%
\bibitem [{\citenamefont {Husmann}\ \emph {et~al.}(2015)\citenamefont {Husmann}, \citenamefont {Uchino}, \citenamefont {Krinner}, \citenamefont {Lebrat}, \citenamefont {Giamarchi}, \citenamefont {Esslinger},\ and\ \citenamefont {Brantut}}]{Husmann_2015}%
  \BibitemOpen
  \bibfield  {author} {\bibinfo {author} {\bibfnamefont {D.}~\bibnamefont {Husmann}}, \bibinfo {author} {\bibfnamefont {S.}~\bibnamefont {Uchino}}, \bibinfo {author} {\bibfnamefont {S.}~\bibnamefont {Krinner}}, \bibinfo {author} {\bibfnamefont {M.}~\bibnamefont {Lebrat}}, \bibinfo {author} {\bibfnamefont {T.}~\bibnamefont {Giamarchi}}, \bibinfo {author} {\bibfnamefont {T.}~\bibnamefont {Esslinger}},\ and\ \bibinfo {author} {\bibfnamefont {J.-P.}\ \bibnamefont {Brantut}},\ }\bibfield  {title} {\bibinfo {title} {Connecting strongly correlated superfluids by a quantum point contact},\ }\href {https://doi.org/10.1126/science.aac9584} {\bibfield  {journal} {\bibinfo  {journal} {Science}\ }\textbf {\bibinfo {volume} {350}},\ \bibinfo {pages} {1498–1501} (\bibinfo {year} {2015})}\BibitemShut {NoStop}%
\bibitem [{\citenamefont {Krinner}\ \emph {et~al.}(2016)\citenamefont {Krinner}, \citenamefont {Lebrat}, \citenamefont {Husmann}, \citenamefont {Grenier}, \citenamefont {Brantut},\ and\ \citenamefont {Esslinger}}]{doi:10.1073/pnas.1601812113}%
  \BibitemOpen
  \bibfield  {author} {\bibinfo {author} {\bibfnamefont {S.}~\bibnamefont {Krinner}}, \bibinfo {author} {\bibfnamefont {M.}~\bibnamefont {Lebrat}}, \bibinfo {author} {\bibfnamefont {D.}~\bibnamefont {Husmann}}, \bibinfo {author} {\bibfnamefont {C.}~\bibnamefont {Grenier}}, \bibinfo {author} {\bibfnamefont {J.-P.}\ \bibnamefont {Brantut}},\ and\ \bibinfo {author} {\bibfnamefont {T.}~\bibnamefont {Esslinger}},\ }\bibfield  {title} {\bibinfo {title} {Mapping out spin and particle conductances in a quantum point contact},\ }\href {https://doi.org/10.1073/pnas.1601812113} {\bibfield  {journal} {\bibinfo  {journal} {Proceedings of the National Academy of Sciences}\ }\textbf {\bibinfo {volume} {113}},\ \bibinfo {pages} {8144} (\bibinfo {year} {2016})}\BibitemShut {NoStop}%
\bibitem [{\citenamefont {Brantut}\ \emph {et~al.}(2013)\citenamefont {Brantut}, \citenamefont {Grenier}, \citenamefont {Meineke}, \citenamefont {Stadler}, \citenamefont {Krinner}, \citenamefont {Kollath}, \citenamefont {Esslinger},\ and\ \citenamefont {Georges}}]{Brantut_2013}%
  \BibitemOpen
  \bibfield  {author} {\bibinfo {author} {\bibfnamefont {J.-P.}\ \bibnamefont {Brantut}}, \bibinfo {author} {\bibfnamefont {C.}~\bibnamefont {Grenier}}, \bibinfo {author} {\bibfnamefont {J.}~\bibnamefont {Meineke}}, \bibinfo {author} {\bibfnamefont {D.}~\bibnamefont {Stadler}}, \bibinfo {author} {\bibfnamefont {S.}~\bibnamefont {Krinner}}, \bibinfo {author} {\bibfnamefont {C.}~\bibnamefont {Kollath}}, \bibinfo {author} {\bibfnamefont {T.}~\bibnamefont {Esslinger}},\ and\ \bibinfo {author} {\bibfnamefont {A.}~\bibnamefont {Georges}},\ }\bibfield  {title} {\bibinfo {title} {A thermoelectric heat engine with ultracold atoms},\ }\href {https://doi.org/10.1126/science.1242308} {\bibfield  {journal} {\bibinfo  {journal} {Science}\ }\textbf {\bibinfo {volume} {342}},\ \bibinfo {pages} {713–715} (\bibinfo {year} {2013})}\BibitemShut {NoStop}%
\bibitem [{\citenamefont {Fabritius}\ \emph {et~al.}(2024)\citenamefont {Fabritius}, \citenamefont {Mohan}, \citenamefont {Talebi}, \citenamefont {Wili}, \citenamefont {Zwerger}, \citenamefont {Huang},\ and\ \citenamefont {Esslinger}}]{Fabritius_2024}%
  \BibitemOpen
  \bibfield  {author} {\bibinfo {author} {\bibfnamefont {P.}~\bibnamefont {Fabritius}}, \bibinfo {author} {\bibfnamefont {J.}~\bibnamefont {Mohan}}, \bibinfo {author} {\bibfnamefont {M.}~\bibnamefont {Talebi}}, \bibinfo {author} {\bibfnamefont {S.}~\bibnamefont {Wili}}, \bibinfo {author} {\bibfnamefont {W.}~\bibnamefont {Zwerger}}, \bibinfo {author} {\bibfnamefont {M.-Z.}\ \bibnamefont {Huang}},\ and\ \bibinfo {author} {\bibfnamefont {T.}~\bibnamefont {Esslinger}},\ }\bibfield  {title} {\bibinfo {title} {Irreversible entropy transport enhanced by fermionic superfluidity},\ }\href {https://doi.org/10.1038/s41567-024-02483-3} {\bibfield  {journal} {\bibinfo  {journal} {Nature Physics}\ }\textbf {\bibinfo {volume} {20}},\ \bibinfo {pages} {1091–1096} (\bibinfo {year} {2024})}\BibitemShut {NoStop}%
\bibitem [{\citenamefont {Corman}\ \emph {et~al.}(2019)\citenamefont {Corman}, \citenamefont {Fabritius}, \citenamefont {H\"ausler}, \citenamefont {Mohan}, \citenamefont {Dogra}, \citenamefont {Husmann}, \citenamefont {Lebrat},\ and\ \citenamefont {Esslinger}}]{Corman_2019}%
  \BibitemOpen
  \bibfield  {author} {\bibinfo {author} {\bibfnamefont {L.}~\bibnamefont {Corman}}, \bibinfo {author} {\bibfnamefont {P.}~\bibnamefont {Fabritius}}, \bibinfo {author} {\bibfnamefont {S.}~\bibnamefont {H\"ausler}}, \bibinfo {author} {\bibfnamefont {J.}~\bibnamefont {Mohan}}, \bibinfo {author} {\bibfnamefont {L.~H.}\ \bibnamefont {Dogra}}, \bibinfo {author} {\bibfnamefont {D.}~\bibnamefont {Husmann}}, \bibinfo {author} {\bibfnamefont {M.}~\bibnamefont {Lebrat}},\ and\ \bibinfo {author} {\bibfnamefont {T.}~\bibnamefont {Esslinger}},\ }\bibfield  {title} {\bibinfo {title} {Quantized conductance through a dissipative atomic point contact},\ }\href {https://doi.org/10.1103/PhysRevA.100.053605} {\bibfield  {journal} {\bibinfo  {journal} {Phys. Rev. A}\ }\textbf {\bibinfo {volume} {100}},\ \bibinfo {pages} {053605} (\bibinfo {year} {2019})}\BibitemShut {NoStop}%
\bibitem [{\citenamefont {Mohan}\ \emph {et~al.}(2024)\citenamefont {Mohan}, \citenamefont {Fabritius}, \citenamefont {Talebi}, \citenamefont {Wili}, \citenamefont {Huang},\ and\ \citenamefont {Esslinger}}]{mohan2024universalentropytransportfar}%
  \BibitemOpen
  \bibfield  {author} {\bibinfo {author} {\bibfnamefont {J.}~\bibnamefont {Mohan}}, \bibinfo {author} {\bibfnamefont {P.}~\bibnamefont {Fabritius}}, \bibinfo {author} {\bibfnamefont {M.}~\bibnamefont {Talebi}}, \bibinfo {author} {\bibfnamefont {S.}~\bibnamefont {Wili}}, \bibinfo {author} {\bibfnamefont {M.-Z.}\ \bibnamefont {Huang}},\ and\ \bibinfo {author} {\bibfnamefont {T.}~\bibnamefont {Esslinger}},\ }\href@noop {} {\bibinfo {title} {Universal entropy transport far from equilibrium across the {BCS-BEC} crossover}} (\bibinfo {year} {2024}),\ \Eprint {https://arxiv.org/abs/2403.17838} {arXiv:2403.17838 [cond-mat.quant-gas]} \BibitemShut {NoStop}%
\bibitem [{\citenamefont {Setiawan}\ and\ \citenamefont {Hofmann}(2022)}]{PhysRevResearch.4.043087}%
  \BibitemOpen
  \bibfield  {author} {\bibinfo {author} {\bibfnamefont {F.}~\bibnamefont {Setiawan}}\ and\ \bibinfo {author} {\bibfnamefont {J.}~\bibnamefont {Hofmann}},\ }\bibfield  {title} {\bibinfo {title} {Analytic approach to transport in superconducting junctions with arbitrary carrier density},\ }\href {https://doi.org/10.1103/PhysRevResearch.4.043087} {\bibfield  {journal} {\bibinfo  {journal} {Phys. Rev. Res.}\ }\textbf {\bibinfo {volume} {4}},\ \bibinfo {pages} {043087} (\bibinfo {year} {2022})}\BibitemShut {NoStop}%
\bibitem [{\citenamefont {Virtanen}\ and\ \citenamefont {Giazotto}(2014)}]{PhysRevB.90.014511}%
  \BibitemOpen
  \bibfield  {author} {\bibinfo {author} {\bibfnamefont {P.}~\bibnamefont {Virtanen}}\ and\ \bibinfo {author} {\bibfnamefont {F.}~\bibnamefont {Giazotto}},\ }\bibfield  {title} {\bibinfo {title} {Thermal transport through ac-driven transparent josephson weak links},\ }\href {https://doi.org/10.1103/PhysRevB.90.014511} {\bibfield  {journal} {\bibinfo  {journal} {Phys. Rev. B}\ }\textbf {\bibinfo {volume} {90}},\ \bibinfo {pages} {014511} (\bibinfo {year} {2014})}\BibitemShut {NoStop}%
\bibitem [{\citenamefont {Plastovets}(2026)}]{Plastovets_2026}%
  \BibitemOpen
  \bibfield  {author} {\bibinfo {author} {\bibfnamefont {V.}~\bibnamefont {Plastovets}},\ }\bibfield  {title} {\bibinfo {title} {Fluctuation response of a superconductor with temporally correlated noise},\ }\href {https://doi.org/10.1088/1361-648x/ae8233} {\bibfield  {journal} {\bibinfo  {journal} {Journal of Physics: Condensed Matter}\ }\textbf {\bibinfo {volume} {38}},\ \bibinfo {pages} {275604} (\bibinfo {year} {2026})}\BibitemShut {NoStop}%
\bibitem [{\citenamefont {Cuevas}\ \emph {et~al.}(1996)\citenamefont {Cuevas}, \citenamefont {Martín-Rodero},\ and\ \citenamefont {Yeyati}}]{Cuevas_1996}%
  \BibitemOpen
  \bibfield  {author} {\bibinfo {author} {\bibfnamefont {J.~C.}\ \bibnamefont {Cuevas}}, \bibinfo {author} {\bibfnamefont {A.}~\bibnamefont {Martín-Rodero}},\ and\ \bibinfo {author} {\bibfnamefont {A.~L.}\ \bibnamefont {Yeyati}},\ }\bibfield  {title} {\bibinfo {title} {Hamiltonian approach to the transport properties of superconducting quantum point contacts},\ }\href {https://doi.org/10.1103/physrevb.54.7366} {\bibfield  {journal} {\bibinfo  {journal} {Physical Review B}\ }\textbf {\bibinfo {volume} {54}},\ \bibinfo {pages} {7366–7379} (\bibinfo {year} {1996})}\BibitemShut {NoStop}%
\bibitem [{\citenamefont {Klapwijk}\ \emph {et~al.}(1982)\citenamefont {Klapwijk}, \citenamefont {Blonder},\ and\ \citenamefont {Tinkham}}]{klapwijk1982}%
  \BibitemOpen
  \bibfield  {author} {\bibinfo {author} {\bibfnamefont {T.~M.}\ \bibnamefont {Klapwijk}}, \bibinfo {author} {\bibfnamefont {G.~E.}\ \bibnamefont {Blonder}},\ and\ \bibinfo {author} {\bibfnamefont {M.}~\bibnamefont {Tinkham}},\ }\bibfield  {title} {\bibinfo {title} {Explanation of subharmonic energy gap structure in superconducting contacts},\ }\href {https://doi.org/10.1016/0378-4363(82)90528-9} {\bibfield  {journal} {\bibinfo  {journal} {Physica B+C}\ }\textbf {\bibinfo {volume} {109--110}} (\bibinfo {year} {1982})}\BibitemShut {NoStop}%
\bibitem [{\citenamefont {Cuevas}\ and\ \citenamefont {Belzig}(2003)}]{Cuevas_2003}%
  \BibitemOpen
  \bibfield  {author} {\bibinfo {author} {\bibfnamefont {J.~C.}\ \bibnamefont {Cuevas}}\ and\ \bibinfo {author} {\bibfnamefont {W.}~\bibnamefont {Belzig}},\ }\bibfield  {title} {\bibinfo {title} {Full counting statistics of multiple {A}ndreev reflections},\ }\href {https://doi.org/10.1103/PhysRevLett.91.187001} {\bibfield  {journal} {\bibinfo  {journal} {Phys. Rev. Lett.}\ }\textbf {\bibinfo {volume} {91}},\ \bibinfo {pages} {187001} (\bibinfo {year} {2003})}\BibitemShut {NoStop}%
\bibitem [{\citenamefont {Bolech}\ and\ \citenamefont {Giamarchi}(2005)}]{Bolech_2005}%
  \BibitemOpen
  \bibfield  {author} {\bibinfo {author} {\bibfnamefont {C.~J.}\ \bibnamefont {Bolech}}\ and\ \bibinfo {author} {\bibfnamefont {T.}~\bibnamefont {Giamarchi}},\ }\bibfield  {title} {\bibinfo {title} {Keldysh study of point-contact tunneling between superconductors},\ }\href {https://doi.org/10.1103/PhysRevB.71.024517} {\bibfield  {journal} {\bibinfo  {journal} {Phys. Rev. B}\ }\textbf {\bibinfo {volume} {71}},\ \bibinfo {pages} {024517} (\bibinfo {year} {2005})}\BibitemShut {NoStop}%
\bibitem [{\citenamefont {Visuri}\ \emph {et~al.}(2023)\citenamefont {Visuri}, \citenamefont {Mohan}, \citenamefont {Uchino}, \citenamefont {Huang}, \citenamefont {Esslinger},\ and\ \citenamefont {Giamarchi}}]{Visuri_2023}%
  \BibitemOpen
  \bibfield  {author} {\bibinfo {author} {\bibfnamefont {A.-M.}\ \bibnamefont {Visuri}}, \bibinfo {author} {\bibfnamefont {J.}~\bibnamefont {Mohan}}, \bibinfo {author} {\bibfnamefont {S.}~\bibnamefont {Uchino}}, \bibinfo {author} {\bibfnamefont {M.-Z.}\ \bibnamefont {Huang}}, \bibinfo {author} {\bibfnamefont {T.}~\bibnamefont {Esslinger}},\ and\ \bibinfo {author} {\bibfnamefont {T.}~\bibnamefont {Giamarchi}},\ }\bibfield  {title} {\bibinfo {title} {{DC} transport in a dissipative superconducting quantum point contact},\ }\href {https://doi.org/10.1103/PhysRevResearch.5.033095} {\bibfield  {journal} {\bibinfo  {journal} {Phys. Rev. Res.}\ }\textbf {\bibinfo {volume} {5}},\ \bibinfo {pages} {033095} (\bibinfo {year} {2023})}\BibitemShut {NoStop}%
\bibitem [{\citenamefont {Levy~Yeyati}\ \emph {et~al.}(1995)\citenamefont {Levy~Yeyati}, \citenamefont {Martín-Rodero},\ and\ \citenamefont {García-Vidal}}]{Levy_Yeyati_1995}%
  \BibitemOpen
  \bibfield  {author} {\bibinfo {author} {\bibfnamefont {A.}~\bibnamefont {Levy~Yeyati}}, \bibinfo {author} {\bibfnamefont {A.}~\bibnamefont {Martín-Rodero}},\ and\ \bibinfo {author} {\bibfnamefont {F.~J.}\ \bibnamefont {García-Vidal}},\ }\bibfield  {title} {\bibinfo {title} {Self-consistent theory of superconducting mesoscopic weak links},\ }\href {https://doi.org/10.1103/physrevb.51.3743} {\bibfield  {journal} {\bibinfo  {journal} {Physical Review B}\ }\textbf {\bibinfo {volume} {51}},\ \bibinfo {pages} {3743–3753} (\bibinfo {year} {1995})}\BibitemShut {NoStop}%
\bibitem [{\citenamefont {Shankar}(1994)}]{shankar1994principles}%
  \BibitemOpen
  \bibfield  {author} {\bibinfo {author} {\bibfnamefont {R.}~\bibnamefont {Shankar}},\ }\href {https://link.springer.com/book/9780306447907} {\emph {\bibinfo {title} {Principles of Quantum Mechanics}}},\ \bibinfo {edition} {2nd}\ ed.\ (\bibinfo  {publisher} {Springer},\ \bibinfo {address} {New York},\ \bibinfo {year} {1994})\BibitemShut {NoStop}%
\bibitem [{\citenamefont {Uchino}(2021)}]{Uchino_2021}%
  \BibitemOpen
  \bibfield  {author} {\bibinfo {author} {\bibfnamefont {S.}~\bibnamefont {Uchino}},\ }\bibfield  {title} {\bibinfo {title} {Tunneling {H}amiltonian analysis of {DC} {J}osephson currents in a weakly-interacting {B}ose-{E}instein condensate},\ }\href {https://doi.org/10.1103/PhysRevResearch.3.043058} {\bibfield  {journal} {\bibinfo  {journal} {Phys. Rev. Res.}\ }\textbf {\bibinfo {volume} {3}},\ \bibinfo {pages} {043058} (\bibinfo {year} {2021})}\BibitemShut {NoStop}%
\bibitem [{\citenamefont {Kamenev}(2011)}]{Kamenev_2011}%
  \BibitemOpen
  \bibfield  {author} {\bibinfo {author} {\bibfnamefont {A.}~\bibnamefont {Kamenev}},\ }\href {https://www.cambridge.org/core/books/field-theory-of-nonequilibrium-systems/CEA995D5C5C7E043E9BAAE6DCA282354} {\emph {\bibinfo {title} {Field Theory of Non-Equilibrium Systems}}}\ (\bibinfo  {publisher} {Cambridge University Press},\ \bibinfo {year} {2011})\BibitemShut {NoStop}%
\bibitem [{\citenamefont {Sieberer}\ \emph {et~al.}(2016)\citenamefont {Sieberer}, \citenamefont {Buchhold},\ and\ \citenamefont {Diehl}}]{Sieberer_2016}%
  \BibitemOpen
  \bibfield  {author} {\bibinfo {author} {\bibfnamefont {L.~M.}\ \bibnamefont {Sieberer}}, \bibinfo {author} {\bibfnamefont {M.}~\bibnamefont {Buchhold}},\ and\ \bibinfo {author} {\bibfnamefont {S.}~\bibnamefont {Diehl}},\ }\bibfield  {title} {\bibinfo {title} {Keldysh field theory for driven open quantum systems},\ }\href {https://doi.org/10.1088/0034-4885/79/9/096001} {\bibfield  {journal} {\bibinfo  {journal} {Reports on Progress in Physics}\ }\textbf {\bibinfo {volume} {79}},\ \bibinfo {pages} {096001} (\bibinfo {year} {2016})}\BibitemShut {NoStop}%
\bibitem [{\citenamefont {Landauer}(1957)}]{Landauer_1957}%
  \BibitemOpen
  \bibfield  {author} {\bibinfo {author} {\bibfnamefont {R.}~\bibnamefont {Landauer}},\ }\bibfield  {title} {\bibinfo {title} {Spatial variation of currents and fields due to localized scatterers in metallic conduction},\ }\href {https://doi.org/10.1147/rd.13.0223} {\bibfield  {journal} {\bibinfo  {journal} {IBM Journal of Research and Development}\ }\textbf {\bibinfo {volume} {1}},\ \bibinfo {pages} {223} (\bibinfo {year} {1957})}\BibitemShut {NoStop}%
\bibitem [{\citenamefont {B\"uttiker}(1986)}]{Buttiker}%
  \BibitemOpen
  \bibfield  {author} {\bibinfo {author} {\bibfnamefont {M.}~\bibnamefont {B\"uttiker}},\ }\bibfield  {title} {\bibinfo {title} {Four-terminal phase-coherent conductance},\ }\href {https://doi.org/10.1103/PhysRevLett.57.1761} {\bibfield  {journal} {\bibinfo  {journal} {Phys. Rev. Lett.}\ }\textbf {\bibinfo {volume} {57}},\ \bibinfo {pages} {1761} (\bibinfo {year} {1986})}\BibitemShut {NoStop}%
\bibitem [{\citenamefont {Berthod}\ and\ \citenamefont {Giamarchi}(2011)}]{Berthod-PRB-2011}%
  \BibitemOpen
  \bibfield  {author} {\bibinfo {author} {\bibfnamefont {C.}~\bibnamefont {Berthod}}\ and\ \bibinfo {author} {\bibfnamefont {T.}~\bibnamefont {Giamarchi}},\ }\bibfield  {title} {\bibinfo {title} {Tunneling conductance and local density of states in tight-binding junctions},\ }\href {http://dx.doi.org/10.1103/PhysRevB.84.155414} {\bibfield  {journal} {\bibinfo  {journal} {Physical Review B}\ }\textbf {\bibinfo {volume} {84}},\ \bibinfo {pages} {155414} (\bibinfo {year} {2011})}\BibitemShut {NoStop}%
\bibitem [{\citenamefont {Todorov}\ \emph {et~al.}(1993)\citenamefont {Todorov}, \citenamefont {Briggs},\ and\ \citenamefont {Sutton}}]{Todorov_1993}%
  \BibitemOpen
  \bibfield  {author} {\bibinfo {author} {\bibfnamefont {T.~N.}\ \bibnamefont {Todorov}}, \bibinfo {author} {\bibfnamefont {G.~A.~D.}\ \bibnamefont {Briggs}},\ and\ \bibinfo {author} {\bibfnamefont {A.~P.}\ \bibnamefont {Sutton}},\ }\bibfield  {title} {\bibinfo {title} {Elastic quantum transport through small structures},\ }\href {https://doi.org/10.1088/0953-8984/5/15/010} {\bibfield  {journal} {\bibinfo  {journal} {Journal of Physics: Condensed Matter}\ }\textbf {\bibinfo {volume} {5}},\ \bibinfo {pages} {2389} (\bibinfo {year} {1993})}\BibitemShut {NoStop}%
\bibitem [{\citenamefont {Kittel}(2005)}]{kittel2005introduction}%
  \BibitemOpen
  \bibfield  {author} {\bibinfo {author} {\bibfnamefont {C.}~\bibnamefont {Kittel}},\ }\href {https://www.wiley.com/en-us/Introduction+to+Solid+State+Physics%2C+8th+Edition-p-9780471415268} {\emph {\bibinfo {title} {Introduction to Solid State Physics}}},\ \bibinfo {edition} {8th}\ ed.\ (\bibinfo  {publisher} {John Wiley \& Sons},\ \bibinfo {address} {Hoboken, NJ},\ \bibinfo {year} {2005})\BibitemShut {NoStop}%
\bibitem [{\citenamefont {Bolech}\ and\ \citenamefont {Giamarchi}(2004)}]{Bolech_triplet}%
  \BibitemOpen
  \bibfield  {author} {\bibinfo {author} {\bibfnamefont {C.~J.}\ \bibnamefont {Bolech}}\ and\ \bibinfo {author} {\bibfnamefont {T.}~\bibnamefont {Giamarchi}},\ }\bibfield  {title} {\bibinfo {title} {Point-contact tunneling involving low-dimensional spin-triplet superconductors},\ }\href {https://doi.org/10.1103/PhysRevLett.92.127001} {\bibfield  {journal} {\bibinfo  {journal} {Phys. Rev. Lett.}\ }\textbf {\bibinfo {volume} {92}},\ \bibinfo {pages} {127001} (\bibinfo {year} {2004})}\BibitemShut {NoStop}%
\bibitem [{\citenamefont {Blonder}\ \emph {et~al.}(2 04)\citenamefont {Blonder}, \citenamefont {Tinkham},\ and\ \citenamefont {Klapwijk}}]{blonder1982}%
  \BibitemOpen
  \bibfield  {author} {\bibinfo {author} {\bibfnamefont {G.~E.}\ \bibnamefont {Blonder}}, \bibinfo {author} {\bibfnamefont {M.}~\bibnamefont {Tinkham}},\ and\ \bibinfo {author} {\bibfnamefont {T.~M.}\ \bibnamefont {Klapwijk}},\ }\bibfield  {title} {\bibinfo {title} {Transition from metallic to tunneling regimes in superconducting microconstrictions: Excess current, charge imbalance, and supercurrent conversion},\ }\href {https://doi.org/10.1103/PhysRevB.25.4515} {\bibfield  {journal} {\bibinfo  {journal} {Phys. Rev. B}\ }\textbf {\bibinfo {volume} {25}},\ \bibinfo {pages} {4515} (\bibinfo {year} {1982-04})}\BibitemShut {NoStop}%
\bibitem [{\citenamefont {Yao}\ \emph {et~al.}(2018)\citenamefont {Yao}, \citenamefont {Liu}, \citenamefont {Sun},\ and\ \citenamefont {Zhai}}]{Yao_2018}%
  \BibitemOpen
  \bibfield  {author} {\bibinfo {author} {\bibfnamefont {J.}~\bibnamefont {Yao}}, \bibinfo {author} {\bibfnamefont {B.}~\bibnamefont {Liu}}, \bibinfo {author} {\bibfnamefont {M.}~\bibnamefont {Sun}},\ and\ \bibinfo {author} {\bibfnamefont {H.}~\bibnamefont {Zhai}},\ }\bibfield  {title} {\bibinfo {title} {Controlled transport between {F}ermi superfluids through a quantum point contact},\ }\href {https://doi.org/10.1103/PhysRevA.98.041601} {\bibfield  {journal} {\bibinfo  {journal} {Phys. Rev. A}\ }\textbf {\bibinfo {volume} {98}},\ \bibinfo {pages} {041601(R)} (\bibinfo {year} {2018})}\BibitemShut {NoStop}%
\bibitem [{Note1()}]{Note1}%
  \BibitemOpen
  \bibinfo {note} {When transport mediated by Cooper pairs created inside the gap is fully suppressed, one can identify a soft upper bound for the eMAR and oMAR processes. After this point, their contribution to the current is reduced, especially in the ballistic limit. These bounds are given by $ \Delta \mu = \protect \frac {2\Delta }{2n+2} \hskip 1em\relax \protect \text {and} \hskip 1em\relax \Delta \mu = \protect \frac {2\Delta }{2n+3},$ for eMAR and oMAR processes, respectively. In particular, for a $2n$-eMAR ($(2n+1)$-oMAR) process this corresponds to the activation of the $(2n+2)$-eMAR ($(2n+3)$-oMAR) process.}\BibitemShut {Stop}%
\bibitem [{Note2()}]{Note2}%
  \BibitemOpen
  \bibinfo {note} {A particularly relevant case is that of a single Andreev reflection. In this regime, the transport window for Cooper pairs within the superconducting gap is not reduced, and persists even at $\Delta \mu \gg \Delta $.}\BibitemShut {Stop}%
\bibitem [{\citenamefont {Mintchev}\ \emph {et~al.}(2017)\citenamefont {Mintchev}, \citenamefont {Santoni},\ and\ \citenamefont {Sorba}}]{PhysRevE.96.052124}%
  \BibitemOpen
  \bibfield  {author} {\bibinfo {author} {\bibfnamefont {M.}~\bibnamefont {Mintchev}}, \bibinfo {author} {\bibfnamefont {L.}~\bibnamefont {Santoni}},\ and\ \bibinfo {author} {\bibfnamefont {P.}~\bibnamefont {Sorba}},\ }\bibfield  {title} {\bibinfo {title} {Quantum fluctuations of entropy production for fermionic systems in the landauer-b\"uttiker state},\ }\href {https://doi.org/10.1103/PhysRevE.96.052124} {\bibfield  {journal} {\bibinfo  {journal} {Phys. Rev. E}\ }\textbf {\bibinfo {volume} {96}},\ \bibinfo {pages} {052124} (\bibinfo {year} {2017})}\BibitemShut {NoStop}%
\bibitem [{\citenamefont {Mintchev}\ and\ \citenamefont {Sorba}(2020)}]{https://doi.org/10.1002/andp.202000276}%
  \BibitemOpen
  \bibfield  {author} {\bibinfo {author} {\bibfnamefont {M.}~\bibnamefont {Mintchev}}\ and\ \bibinfo {author} {\bibfnamefont {P.}~\bibnamefont {Sorba}},\ }\bibfield  {title} {\bibinfo {title} {Anyon quantum transport and noise away from equilibrium},\ }\href {https://doi.org/https://doi.org/10.1002/andp.202000276} {\bibfield  {journal} {\bibinfo  {journal} {Annalen der Physik}\ }\textbf {\bibinfo {volume} {532}},\ \bibinfo {pages} {2000276} (\bibinfo {year} {2020})}\BibitemShut {NoStop}%
\bibitem [{\citenamefont {Landi}\ and\ \citenamefont {Paternostro}(2021)}]{RevModPhys.93.035008}%
  \BibitemOpen
  \bibfield  {author} {\bibinfo {author} {\bibfnamefont {G.~T.}\ \bibnamefont {Landi}}\ and\ \bibinfo {author} {\bibfnamefont {M.}~\bibnamefont {Paternostro}},\ }\bibfield  {title} {\bibinfo {title} {Irreversible entropy production: From classical to quantum},\ }\href {https://doi.org/10.1103/RevModPhys.93.035008} {\bibfield  {journal} {\bibinfo  {journal} {Rev. Mod. Phys.}\ }\textbf {\bibinfo {volume} {93}},\ \bibinfo {pages} {035008} (\bibinfo {year} {2021})}\BibitemShut {NoStop}%
\bibitem [{\citenamefont {Child}\ \emph {et~al.}(2022)\citenamefont {Child}, \citenamefont {Sheekey}, \citenamefont {L\"uscher}, \citenamefont {Fallahi}, \citenamefont {Gardner}, \citenamefont {Manfra}, \citenamefont {Mitchell}, \citenamefont {Sela}, \citenamefont {Kleeorin}, \citenamefont {Meir},\ and\ \citenamefont {Folk}}]{PhysRevLett.129.227702}%
  \BibitemOpen
  \bibfield  {author} {\bibinfo {author} {\bibfnamefont {T.}~\bibnamefont {Child}}, \bibinfo {author} {\bibfnamefont {O.}~\bibnamefont {Sheekey}}, \bibinfo {author} {\bibfnamefont {S.}~\bibnamefont {L\"uscher}}, \bibinfo {author} {\bibfnamefont {S.}~\bibnamefont {Fallahi}}, \bibinfo {author} {\bibfnamefont {G.~C.}\ \bibnamefont {Gardner}}, \bibinfo {author} {\bibfnamefont {M.}~\bibnamefont {Manfra}}, \bibinfo {author} {\bibfnamefont {A.}~\bibnamefont {Mitchell}}, \bibinfo {author} {\bibfnamefont {E.}~\bibnamefont {Sela}}, \bibinfo {author} {\bibfnamefont {Y.}~\bibnamefont {Kleeorin}}, \bibinfo {author} {\bibfnamefont {Y.}~\bibnamefont {Meir}},\ and\ \bibinfo {author} {\bibfnamefont {J.}~\bibnamefont {Folk}},\ }\bibfield  {title} {\bibinfo {title} {Entropy measurement of a strongly coupled quantum dot},\ }\href {https://doi.org/10.1103/PhysRevLett.129.227702} {\bibfield  {journal} {\bibinfo  {journal} {Phys. Rev. Lett.}\ }\textbf {\bibinfo {volume} {129}},\ \bibinfo {pages} {227702} (\bibinfo {year}
  {2022})}\BibitemShut {NoStop}%
\bibitem [{\citenamefont {Hartman}\ \emph {et~al.}(2018)\citenamefont {Hartman}, \citenamefont {Olsen}, \citenamefont {Lüscher} \emph {et~al.}}]{Hartman2018}%
  \BibitemOpen
  \bibfield  {author} {\bibinfo {author} {\bibfnamefont {N.}~\bibnamefont {Hartman}}, \bibinfo {author} {\bibfnamefont {C.}~\bibnamefont {Olsen}}, \bibinfo {author} {\bibfnamefont {S.}~\bibnamefont {Lüscher}}, \emph {et~al.},\ }\bibfield  {title} {\bibinfo {title} {Direct entropy measurement in a mesoscopic quantum system},\ }\href {https://doi.org/10.1038/s41567-018-0250-5} {\bibfield  {journal} {\bibinfo  {journal} {Nature Physics}\ }\textbf {\bibinfo {volume} {14}},\ \bibinfo {pages} {1083} (\bibinfo {year} {2018})}\BibitemShut {NoStop}%
\bibitem [{\citenamefont {Pyurbeeva}\ \emph {et~al.}(2021)\citenamefont {Pyurbeeva}, \citenamefont {Hsu}, \citenamefont {Vogel}, \citenamefont {Wegeberg}, \citenamefont {Mayor}, \citenamefont {van~der Zant}, \citenamefont {Mol},\ and\ \citenamefont {Gehring}}]{doi:10.1021/acs.nanolett.1c03591}%
  \BibitemOpen
  \bibfield  {author} {\bibinfo {author} {\bibfnamefont {E.}~\bibnamefont {Pyurbeeva}}, \bibinfo {author} {\bibfnamefont {C.}~\bibnamefont {Hsu}}, \bibinfo {author} {\bibfnamefont {D.}~\bibnamefont {Vogel}}, \bibinfo {author} {\bibfnamefont {C.}~\bibnamefont {Wegeberg}}, \bibinfo {author} {\bibfnamefont {M.}~\bibnamefont {Mayor}}, \bibinfo {author} {\bibfnamefont {H.}~\bibnamefont {van~der Zant}}, \bibinfo {author} {\bibfnamefont {J.~A.}\ \bibnamefont {Mol}},\ and\ \bibinfo {author} {\bibfnamefont {P.}~\bibnamefont {Gehring}},\ }\bibfield  {title} {\bibinfo {title} {Controlling the entropy of a single-molecule junction},\ }\href {https://doi.org/10.1021/acs.nanolett.1c03591} {\bibfield  {journal} {\bibinfo  {journal} {Nano Letters}\ }\textbf {\bibinfo {volume} {21}},\ \bibinfo {pages} {9715} (\bibinfo {year} {2021})}\BibitemShut {NoStop}%
\bibitem [{\citenamefont {Chen}(2021)}]{chen2021introduction}%
  \BibitemOpen
  \bibfield  {author} {\bibinfo {author} {\bibfnamefont {C.~J.}\ \bibnamefont {Chen}},\ }\href {https://doi.org/10.1093/oso/9780198856559.001.0001} {\emph {\bibinfo {title} {Introduction to Scanning Tunneling Microscopy}}},\ \bibinfo {edition} {3rd}\ ed.\ (\bibinfo  {publisher} {Oxford University Press},\ \bibinfo {year} {2021})\BibitemShut {NoStop}%
\bibitem [{\citenamefont {Levi}\ \emph {et~al.}(1997)\citenamefont {Levi}, \citenamefont {Millo}, \citenamefont {Rizzo}, \citenamefont {Prober},\ and\ \citenamefont {Motowidlo}}]{Levi1997}%
  \BibitemOpen
  \bibfield  {author} {\bibinfo {author} {\bibfnamefont {Y.}~\bibnamefont {Levi}}, \bibinfo {author} {\bibfnamefont {O.}~\bibnamefont {Millo}}, \bibinfo {author} {\bibfnamefont {N.~D.}\ \bibnamefont {Rizzo}}, \bibinfo {author} {\bibfnamefont {D.~E.}\ \bibnamefont {Prober}},\ and\ \bibinfo {author} {\bibfnamefont {L.~R.}\ \bibnamefont {Motowidlo}},\ }\bibfield  {title} {\bibinfo {title} {{STM} studies of the proximity effect in superconducting wires with artificial pinning centers},\ }\href {https://bibliotekanauki.pl/articles/1968775} {\bibfield  {journal} {\bibinfo  {journal} {Acta Physica Polonica A}\ }\textbf {\bibinfo {volume} {93}} (\bibinfo {year} {1997})},\ \bibinfo {note} {proceedings of the 1st International Symposium on Scanning Probe Spectroscopy and Related Methods, Poznań 1997}\BibitemShut {NoStop}%
\bibitem [{\citenamefont {Bertolusso}\ \emph {et~al.}(2026)\citenamefont {Bertolusso}, \citenamefont {Bolech},\ and\ \citenamefont {Giamarchi}}]{bertolusso_2026_22014928}%
  \BibitemOpen
  \bibfield  {author} {\bibinfo {author} {\bibfnamefont {D.}~\bibnamefont {Bertolusso}}, \bibinfo {author} {\bibfnamefont {C.~J.}\ \bibnamefont {Bolech}},\ and\ \bibinfo {author} {\bibfnamefont {T.}~\bibnamefont {Giamarchi}},\ }\href {https://doi.org/10.5281/zenodo.22014928} {\bibinfo {title} {Entropy transport through a superfluid quantum point contact: A keldysh field-theory approach}} (\bibinfo {year} {2026})\BibitemShut {NoStop}%
\bibitem [{\citenamefont {Mahan}(2000)}]{mahan_many-particle_2000}%
  \BibitemOpen
  \bibfield  {author} {\bibinfo {author} {\bibfnamefont {G.~D.}\ \bibnamefont {Mahan}},\ }\href {https://doi.org/10.1007/978-1-4757-5714-9} {\emph {\bibinfo {title} {Many-Particle Physics}}},\ \bibinfo {edition} {3rd}\ ed.\ (\bibinfo  {publisher} {Springer},\ \bibinfo {address} {New York, NY},\ \bibinfo {year} {2000})\BibitemShut {NoStop}%
\bibitem [{\citenamefont {Mohan}(2024)}]{Mohan2024Universal}%
  \BibitemOpen
  \bibfield  {author} {\bibinfo {author} {\bibfnamefont {J.}~\bibnamefont {Mohan}},\ }\emph {\bibinfo {title} {Universal particle and entropy transport in strongly interacting Fermi gases far from equilibrium}},\ \href {https://doi.org/10.3929/ethz-b-000677320} {\bibinfo {type} {Doctoral thesis}},\ \bibinfo  {school} {ETH Zurich} (\bibinfo {year} {2024})\BibitemShut {NoStop}%
\bibitem [{\citenamefont {Callen}(1985)}]{Callen:450289}%
  \BibitemOpen
  \bibfield  {author} {\bibinfo {author} {\bibfnamefont {H.~B.}\ \bibnamefont {Callen}},\ }\href {https://cds.cern.ch/record/450289} {\emph {\bibinfo {title} {Thermodynamics and an introduction to thermostatistics}}}\ (\bibinfo  {publisher} {Wiley},\ \bibinfo {address} {New York, NY},\ \bibinfo {year} {1985})\BibitemShut {NoStop}%
\bibitem [{\citenamefont {{Piessens}}\ \emph {et~al.}(1983)\citenamefont {{Piessens}}, \citenamefont {{de Doncker-Kapenga}}, \citenamefont {{\"Ueberhuber}},\ and\ \citenamefont {{Kahaner}}}]{1983qspa.book.....P}%
  \BibitemOpen
  \bibfield  {author} {\bibinfo {author} {\bibfnamefont {R.}~\bibnamefont {{Piessens}}}, \bibinfo {author} {\bibfnamefont {E.}~\bibnamefont {{de Doncker-Kapenga}}}, \bibinfo {author} {\bibfnamefont {C.~W.}\ \bibnamefont {{\"Ueberhuber}}},\ and\ \bibinfo {author} {\bibfnamefont {D.~K.}\ \bibnamefont {{Kahaner}}},\ }\href {https://doi.org/10.1007/978-3-642-61786-7} {\emph {\bibinfo {title} {Quadpack: A Subroutine Package for Automatic Integration}}},\ Springer Series in Computational Mathematics\ (\bibinfo  {publisher} {Springer Berlin, Heidelberg},\ \bibinfo {year} {1983})\BibitemShut {NoStop}%
\end{thebibliography}%

\end{document}